\documentclass[10pt,journal,cspaper]{IEEEtran}

\usepackage[ruled]{algorithm2e}
\usepackage[font=small]{caption}
\usepackage{subfigure}
\usepackage{booktabs}
\usepackage{bigstrut}
\usepackage{CJK}
\usepackage{enumerate}
\usepackage{cite}
\usepackage{graphicx}
 \usepackage{epstopdf}
 \graphicspath{{../pdf/}{../jpeg/}}
\usepackage[cmex10]{amsmath}
\usepackage{amsfonts}
\usepackage{amssymb}
\usepackage{bm}
\usepackage{algorithmic}
\usepackage{mdwmath}
\usepackage{multirow}
\usepackage{mdwtab}
\usepackage{array}
\usepackage{tikz}
\usetikzlibrary{positioning}
\usetikzlibrary{shapes,snakes}

\makeatletter

\newcommand{\Rmnum}[1]{\expandafter\@slowromancap\romannumeral #1@}
\makeatother

\newtheorem{definition}{Definition}

\begin{document}

\title{Nonnegative Multi-level Network Factorization for Latent Factor Analysis}

\author{Junyu~Xuan,~Jie Lu,~\IEEEmembership{Senior Member,~IEEE},~Xiangfeng~Luo,~\IEEEmembership{Member,~IEEE}\\and~Guangquan Zhang
\IEEEcompsocitemizethanks{
\IEEEcompsocthanksitem J. Xuan is with the Centre for Quantum Computation and Intelligent Systems (QCIS), School of Software, Faculty of Engineering
and Information Technology, University of Technology, Sydney (UTS), Australia and the School of Computer Engineering and Science, Shanghai University, China (e-mail: Junyu.Xuan@student.uts.edu.au).
\IEEEcompsocthanksitem J. Lu is with the Centre for Quantum Computation and
Intelligent Systems (QCIS), School of Software, Faculty of Engineering
and Information Technology, University of Technology, Sydney (UTS),
Australia (e-mail: Jie.Lu@uts.edu.au).
\IEEEcompsocthanksitem X. Luo is with the School of Computer Engineering and Science, Shanghai University, China. (e-mail: luoxf@shu.edu.cn).
\IEEEcompsocthanksitem G. Zhang is with the Centre for Quantum Computation and
Intelligent Systems (QCIS), School of Software, Faculty of Engineering
and Information Technology, University of Technology, Sydney (UTS),
Australia (e-mail: Guangquan.Zhang@uts.edu.au).
}
\thanks{}}

\markboth{}%
{Xuan \MakeLowercase{\textit{et al.}}: Network-constrained Nonnegative Matrix Factorization for Latent Factor Analysis}

\IEEEcompsoctitleabstractindextext{%
\begin{abstract}

Nonnegative Matrix Factorization (NMF) aims to factorize a matrix into two optimized nonnegative matrices and has been widely used for unsupervised learning tasks such as product recommendation based on a rating matrix. However, although networks between nodes with the same nature exist, standard NMF overlooks them, e.g., the social network between users. This problem leads to comparatively low recommendation accuracy because these networks are also reflections of the nature of the nodes, such as the preferences of users in a social network. Also, social networks, as complex networks, have many different structures. Each structure is a composition of links between nodes and reflects the nature of nodes, so retaining the different network structures will lead to differences in recommendation performance. To investigate the impact of these network structures on the factorization, this paper proposes four multi-level network factorization algorithms based on the standard NMF, which integrates the vertical network (e.g., rating matrix) with the structures of horizontal network (e.g., user social network). These algorithms are carefully designed with corresponding convergence proofs to retain four desired network structures. Experiments on synthetic data show that the proposed algorithms are able to preserve the desired network structures as designed. Experiments on real-world data show that considering the horizontal networks improves the accuracy of document clustering and recommendation with standard NMF, and various structures show their differences in performance on these two tasks. These results can be directly used in document clustering and recommendation systems.

\end{abstract}

}

\maketitle

\IEEEdisplaynotcompsoctitleabstractindextext

\IEEEpeerreviewmaketitle

\section{Introduction}

\begin{figure*}[t]
\begin{minipage}[t]{0.3\linewidth}
\centerline{\includegraphics[scale=0.3]{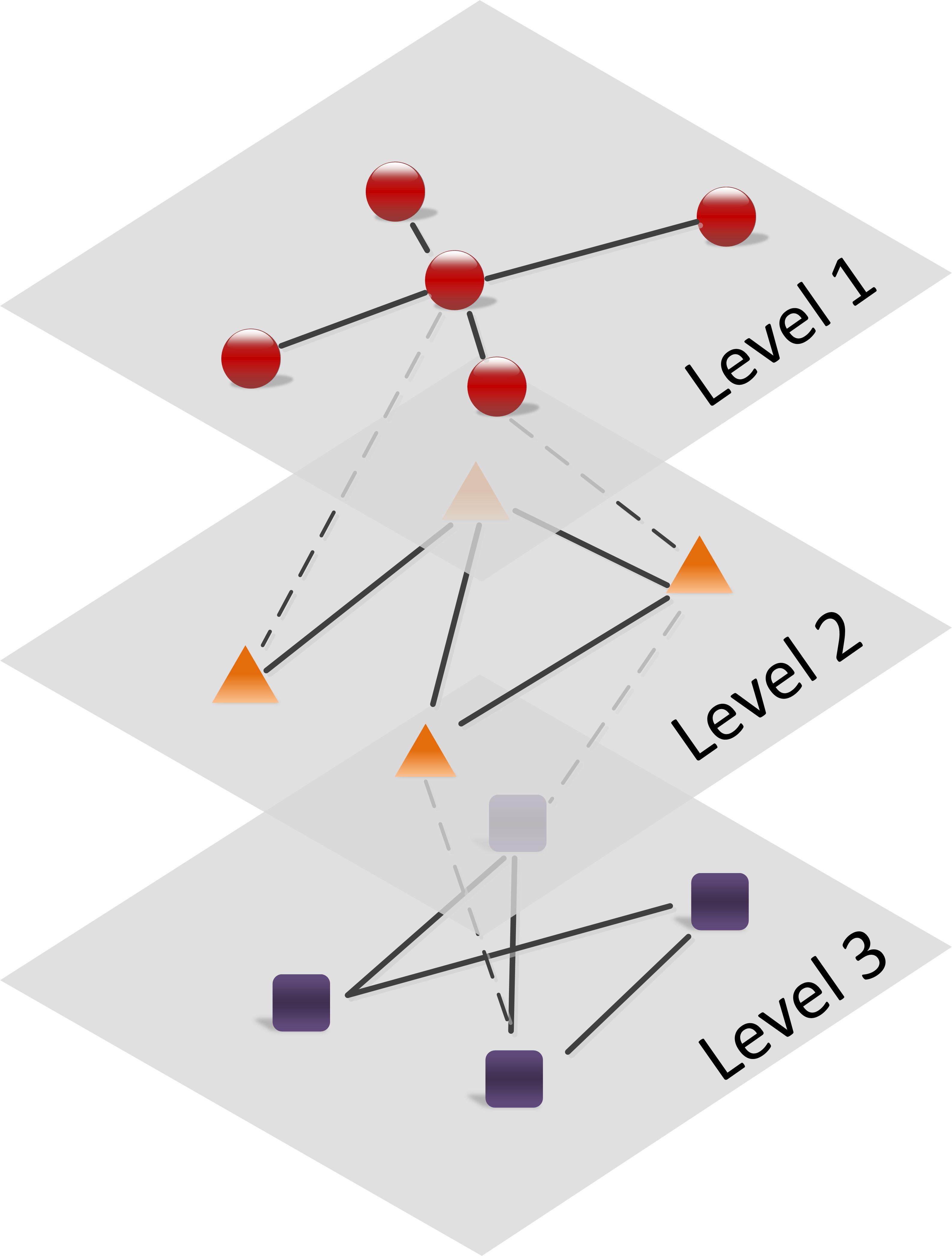}}
\caption{Multi-level network}
\label{mln}
\end{minipage}%
\begin{minipage}[t]{0.35\linewidth}
\centerline{\includegraphics[scale=0.3]{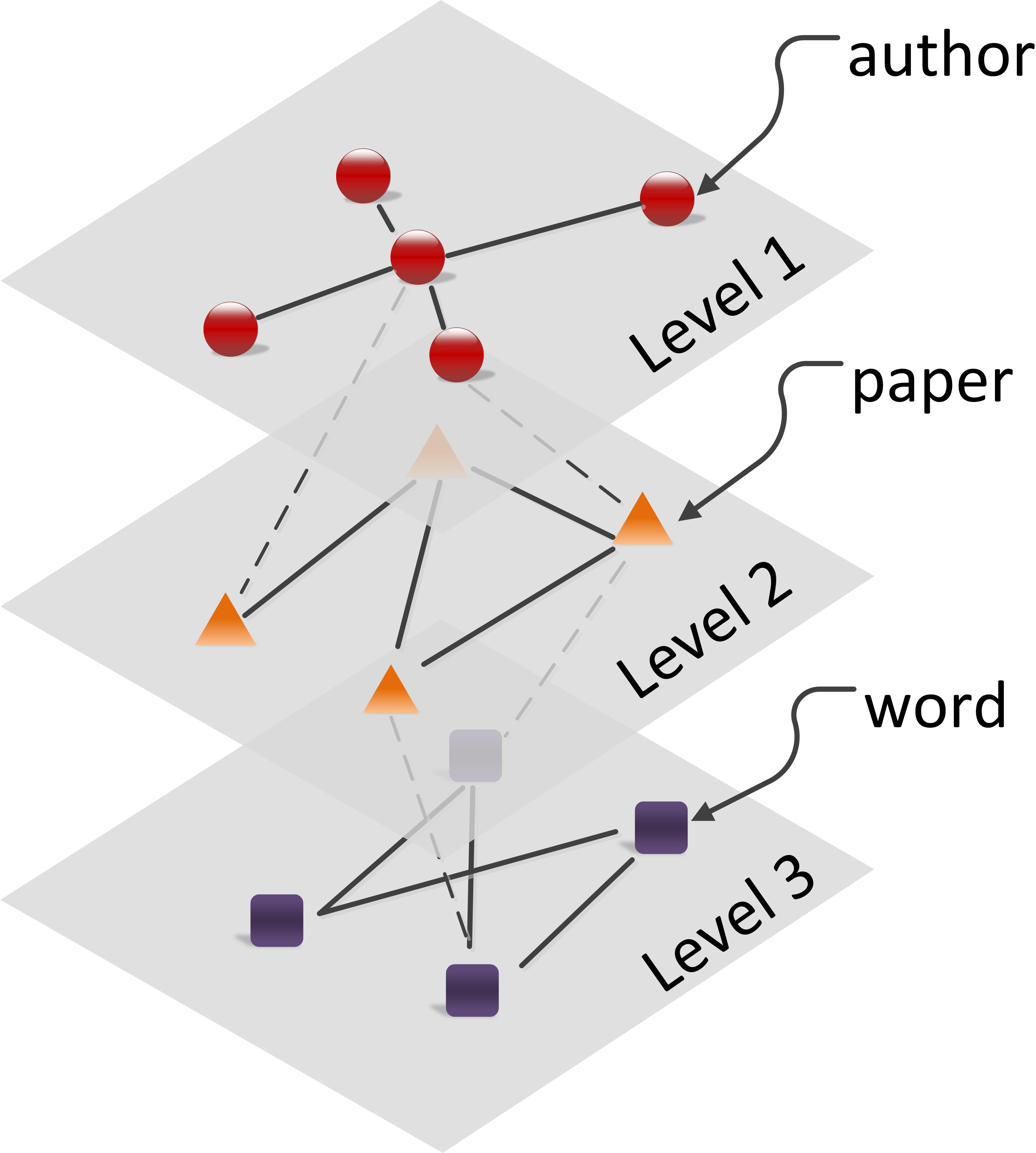}}
\caption{An instance in Text mining}
\label{fig:tm}
\end{minipage}
\begin{minipage}[t]{0.35\linewidth}
\centerline{\includegraphics[scale=0.3]{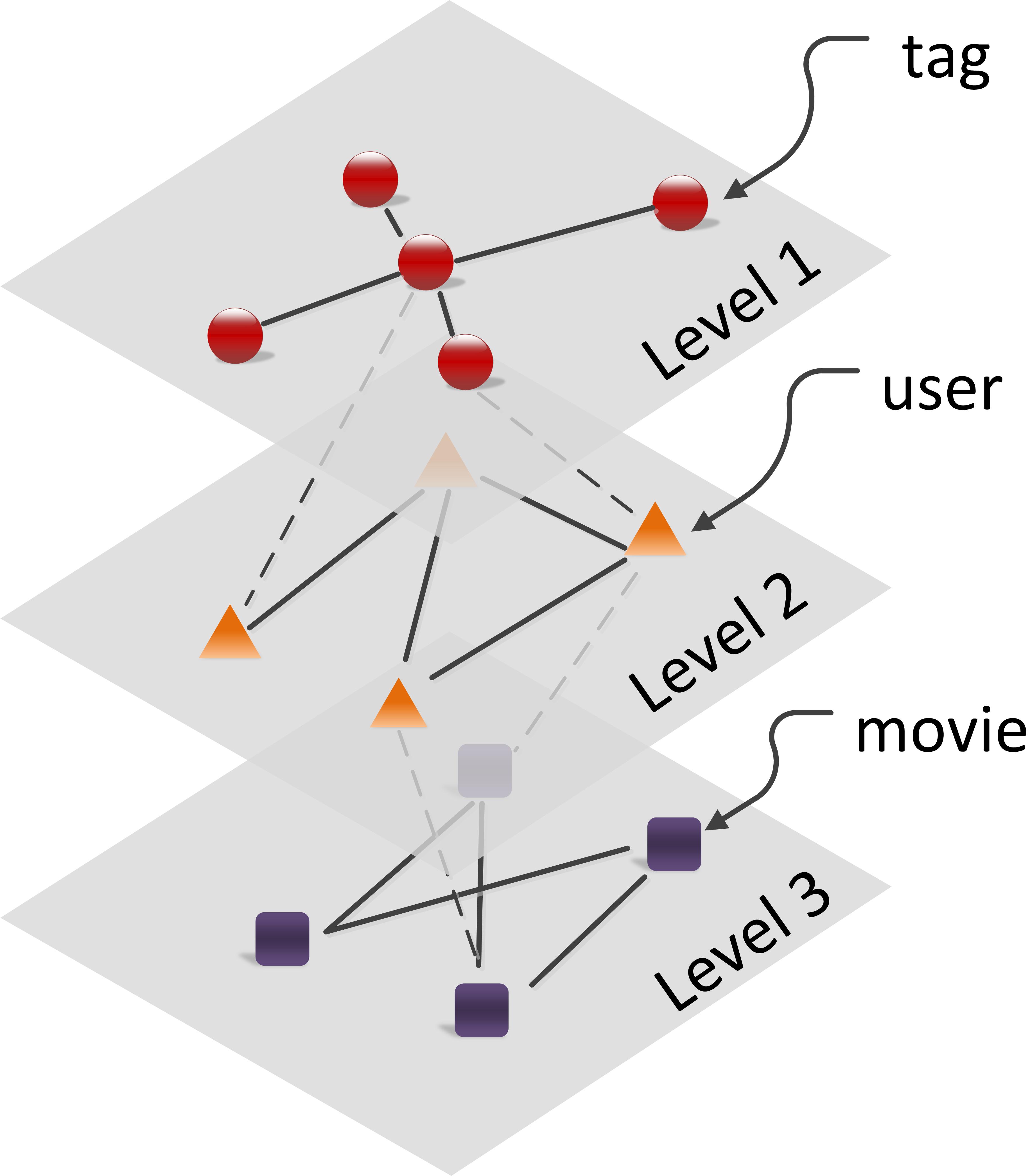}}
\caption{An instance in Recommender Systems}
\label{fig:rs}
\end{minipage}
\end{figure*}

\IEEEPARstart{M}{ulti-level} network is a structure that is composed of vertically connected nodes with different characters and horizontally connected nodes with the same characters. It is an abstract structure, as shown in Fig. \ref{mln}, that can be used to model the data from many areas. One example, in the text mining area, is author-paper-keyword mapping relations (vertical network) with authors' social network, paper citation network and keyword co-occurrence relation network (horizontal networks), as shown in Fig. \ref{fig:tm}. In recommender systems, a multi-level network is composed of tag-user-movie mapping relations (vertical network) with tag similarity network, user social network and movie similarity network (horizontal networks), as shown in Fig. \ref{fig:rs}. The multi-level network structure is very common, especially in a big data environment. Due to the Variety property of big data \cite{hitzler2013linked}, multiple sources and multiple attributes appear to describe the same data. To comprehensively model this kind of data, all the sources and attributes need to be considered simultaneously and so a multi-level network is a good choice. However, it is commonly believed that multi-level networks are sparse and redundant, because the scale of the nodes is normally much larger than the scale of the links.

Considering the sparsity and redundancy of the multi-level network, it would be helpful if an \emph{optimized} latent space for nodes could be found according to its structure. This task is called \emph{factorization}. Take text mining as an example. Suppose we have a multi-level network structure with 1000 authors, 10000 papers, and 8000 keywords. If we find a 100-dimensional space and map all the nodes into this space, we can represent each node, i.e., an author, only by a 100-dimensional vector. The term \emph{optimized} means that the multi-level network structure can be reconstructed by the new representations of nodes with minimum error, such as reconstructing a link between an author and a paper by the cosine similarity between two corresponding 100-dimensional vectors (new representations) for the author and paper. Under this constraint, it is believed that this new representation is not only more concise but also more intrinsic, because the original multi-level network structure as unchanged as possible (some redundant information is removed but the significant information is kept) \cite{6165290,hoyer2004non}. Many real-world applications can benefit from this new concise and intrinsic representation; for example,
\begin{itemize}
  \item \textbf{Document Clustering}: after the factorization of the author-paper-keyword structure, we can cluster papers with similar content together, and then help to organize and retrieve the papers or webpages;
  \item \textbf{Recommendation}: after the factorization of the tag-user-movie structure, we can recommend a movie to a user by the similarity between this movie and user using their concise representations, or recommend a tag to a user in the same fashion.
\end{itemize}
Motivated by these real-world applications, we want to design algorithms in this study that will define and implement multi-level network factorization.

Existing factorization algorithms cannot complete this task, because they overlook the horizontal network structures. The most classical algorithm is Nonnegative Matrix Factorization (NMF)\cite{6165290}. NMF is an elegant tool for factorizing a matrix into two nonnegative matrices, which has been widely used in visual tracking\cite{6579656}, maximum margin classification \cite{6408294}, face recognition \cite{6203594} and probabilistic clustering \cite{6061964}. Two notable inference algorithms to resolve NMF are Alternative Least Square \cite{liu2013solving} and Multiplicative Iterative algorithms \cite{NIPS2000_1861}. Here, we want to interpret NMF on a matrix $M$ as \emph{a bipartite network factorization} so that we can conduct our multi-level network factorization based on it. A matrix $M$ can be seen as a bipartite network (vertical network) with rows as one kind of nodes and columns as another kind of nodes, while each element of matrix $m_{i,j}$ denotes a link between two nodes with different characters. The aim of NMF on a matrix is to find an optimized and relatively small latent space according to the bipartite network structure hidden in this matrix. However, the standard NMF cannot be directly used for the multi-level network because it only considers the vertical network. In fact, the difficulty of multi-level network factorization mainly lies in how to jointly conduct the vertical network factorization and horizontal network factorization.

In this paper, we propose four multi-level network factorization algorithms for preserving four different network structures of the horizonal network, including the whole network structure, community structure, degree distribution structure and max spanning tree structure. These structures are expressed by different link compositions which are sourced from the different natures of nodes. For example, if a user has a range of friends in a social network, it shows that he/she tends to have a range of preferences in movies. In this example, the friend relations of a user express the user's nature (preferences in movies). A natural problem is what is the difference between these network structures in the impact of the recommendation or clustering performance. To evaluate these differences, we conduct experiments to compare the performance of each on two real-world tasks. As the experimental results show, we can achieve better performance by preserving the structures instead of disregarding the horizontal network. We also compare the different performance of retaining different network structures on these tasks.
Four cost functions are carefully designed to preserve the desired network structures. To optimize each cost function, the corresponding update equations are introduced with convergence analysis. Experiments on synthetic data show the designed algorithms have the ability to preserve the desired network structures. To show the usefulness of these algorithms, two real-world tasks, document clustering and recommendation, have been carried out. The results show that the proposed algorithms perform better than traditional NMF in achieving the clustering and recommendation accuracy.

The contributions of this paper are:
\begin{enumerate}
  \item A general multi-level network which can jointly express the relations between nodes that have the same or different natures by the vertical network and horizontal networks is proposed to model different kinds of data;
  \item Based on the nonnegative matrix factorization, four multi-level network factorization algorithms with their convergence proofs are carefully designed to discover the latent space, with different horizontal network structures as constraints.
\end{enumerate}

The rest of this paper is organized as follows. Section \Rmnum{2} reviews related work. The multi-level network and its factorization are formally defined in Section \Rmnum{3}. Our algorithms for multi-level network factorization and their convergence analysis are proposed in Section \Rmnum{4}. Experiments on synthetic data and real-world data are conducted in Section V. Lastly, Section \Rmnum{6} concludes the study and discusses future work.


\section{Related Work}

Since our motivation is to use complex network structures as constraints for nonnegative matrix factorization, this section is composed of two parts: 1) we will discus elementary introductions to, and research on, the structures of a complex network, and 2) we will discus recent works on factorizations with networks.

\subsection{One-level and Multi-level Complex Network}

Complex network is an interdisciplinary research, attracting researchers from computer science, physics, sociology, biology, and so on. Due to the pervasiveness of the network phenomenon, complex network has been adopted to model many things, such as the users' friend network and the cell network in the brain. Comparing to the graph, the complex network area focuses more on the non-trivial structures. The two outstanding structures are small-world network published in Nature \cite{watts1998collective} and power-law degree distribution \cite{Bara15101999} published in Science. In fact, many other different network structures have also been discovered and defined in this area \cite{newman2003structure,estrada2011structure}. However, it is commonly accepted that the following network structures are the most fundamental and significant for describing the structure of a complex network: community structure\cite{6746173,5677532}, degree distribution\cite{hadley2012new,centola2010spread,Adamic24032000}, and max spanning tree\cite{mst,march2010fast}. To the best of our knowledge, little work has been done to consider the influences of complex network structures on factorization.

Recently, the multi-layer/multi-level network has attracted the attention of researchers. Its mathematical formulation is given in  \cite{PhysRevX.3.041022}. Similar to the one-layer complex network, its structures are defined and discussed in \cite{Boccaletti20141}. Apart from formalization and structure definition, the multi-layer network has been used for modeling the influence propagation over microblogs \cite{mlnetwork} and the analysis and management of change propagation \cite{multinetwork2012}. However, most state-of-the-art research of the researches on multi-layer/multi-level networks focuses on basic structure analysis. There is no work on factorization and its applications for document clustering or recommender systems.

Since the traditional network structures (i.e., community, degree distribution and max spanning tree) do not consider the directions of the edges in the network and our aim is to preserve these network structures after the factorization, we assume in this paper that the network is undirected.

\subsection{Network-related factorization models/algorithms}

The existing network-related models/algorithms in recommender systems and document clustering are mainly dominated by two renowned techniques: nonnegative matrix factorization and topic model.

\subsubsection{Network-related Nonnegative Matrix Factorization}

First, we give a brief introduction to nonnegative matrix factorization (NMF). Given a nonnegative matrix $Y_{a \times x}$ (extended to semi-nonnegative by \cite{4685898}), the NMF aims to find two matrices $A_{a \times k}$ and $X_{k \times x}$ to minimize the following cost function
\begin{equation}
\begin{aligned}
J \left ( A, X \right ) =& \frac{1}{2} {\| Y - AX \|}_F^2,
\end{aligned}
\label{nmfcf}
\end{equation}
where $\| \cdot \|_F^2$ is the Frobenius norm and the elements of $A$ and $X$ are also nonnegative.
In the literature, constraints are added to the $A$ or $X$ in the cost function in Eq. (\ref{nmfcf}) for purposes such as sparseness constraint\cite{hoyer2004non}, smooth constraint\cite{6722969}, orthogonal constraint\cite{Li2010905}, and label information \cite{6072214}. All these constraints aim to make the discovered ($k$-dimensional) latent space preserve more properties.

The networks/relations between data have been considered in a number of ways in NMF. One way is to let users define the must-link and cannot-link relations between data \cite{chen2008non}, and then use these relations as constraints for the NMF. Another way is to define a graph between data as the constraint of NMF (also called graph-embedding\cite{huh2013supervised,gu2010} or graph-regularization \cite{5674058,5447650}). As we prove later, this constraint only preserves the community structure of network.
Some works have jointly considered two-side information during factorization. For example, the constraint in \cite{Cui:2011} considers user similarity network and post content; the constraint in \cite{Cheng:2013} considers graphs from multiple domains. There are also works on using NMF to find the community structures of a network \cite{Lin:2011:CDV} or two networks \cite{cmdmkd} but not as constraints, as in this paper.
These works are similar to this paper, but they do not explore the structures of graphs. As we introduced in Section II.A, there are many important structures for a given network. However, these network structures are disregarded during factorization. Our contribution is that the different network structures are considered during factorization.

\subsubsection{Network-related Topic Models}

Topic models \cite{Blei:2012} were originally developed to discover the hidden topics in documents, which can also be seen as a kind of factorization.
Recently, some extensions of the topic models have attempted to adapt the data to  network structures. Since the social network and citation network are two explicit and commonly used networks, most topic models try to adapt to these two types of network. For the social network, the Author-Recipient-Topic (ART) model \cite{McCallum:2007} has been proposed to analyse the categories of roles in the social network, based on the relations of people in the network. A similar task is investigated in \cite{Cha:2012}. The social network is inferred from informal chat-room conversations utilizing the topic model \cite{Tuulos:2004}. The `noisy links' and `popularity bias' of the social network are addressed by properly designed topic models in \cite{5967741} and \cite{Cha:2013:IPT}. As an important issue of social network analysis, communities \cite{Mei:2008:TMN} are extracted by Social Topic Model (STM) \cite{pathak2008social}. Mixed Membership Stochastic Block model is another way to learn the mixed membership vector (i.e., topic distribution) for each node of the community structure \cite{Airoldi:2008:MMS}. For the citation network, Relational Topic Model (RTM) \cite{chang2009relational} is proposed to infer the topics and hierarchical topics from document networks by introducing a link variable between two documents. Unlike RTM, a block is adopted to model the link between two documents \cite{Zhu:2013:STL}. To retain the document network, Markov Random Field (MRF) is combined with topic model \cite{5360275}. The communities in the citation network are also investigated \cite{Liu:2009:TLJ}.

Some models have also considered the hierarchical structure. Original topic model only consider two levels: document level and keyword level. Since there are many features associated with documents, each feature can be seen as the third level in the hierarchical structure, including conference \cite{4781224}, time \cite{Wang:2006:TOT}, author \cite{Rosen-Zvi:2010:LAM}, entity \cite{6413746}, emotion \cite{6007133} and other labels \cite{mcauliffe2008supervised}. Through these specified models, not only can the topic distribution of each document be discovered, but the topic distribution of these features can also be learned. However, all these works disregard the networks. At each level, the items are considered as independent with each other.

To summarize, many researchers have noticed the importance of network structure, and it has been considered in a variety of models. From their work, we can see that the network structure indeed can help to unveil the nature of data. However, other complex network structures (i.e., degree distribution and max spanning tree) are overlooked.


\section{Problem Definition}

In this section, we will formally define and explain multi-level network and its factorization, as well as the problem encountered in multi-level network factorization. The designed algorithms to resolve this problem will be given in the subsequent section.

\subsection{Multi-level Network}

\begin{definition}[Multi-level Network, $\Omega$] Multi-level Network is composed of nodes with different characters. It includes horizontal networks between nodes with the same character and vertical networks between nodes with a different character, as shown in Fig. \ref{mln}.
\end{definition}

This is an abstract and very common model that can be used to model data from different areas. For example,
\begin{itemize}
  \item in recommender systems, there is a multi-level network structure: tag-user-movie, in which users may have trust relations with each other, tags may have correlation relations with each other and movies may have similarity relations due to their genre information;
  \item in the text mining area, there is a multi-level network structure: author-paper-keyword, in which authors may have cooperation relations with each other, papers may have citation relations with each other and keywords may have semantic relations with each other.
\end{itemize}

\subsection{Horizontal-network Factorization}

Given a horizontal network $H_{n \times n}$ in $\Omega$, and the element $h_{n_i, n_j}=w$ denotes that there is an edge between node $n_i$ and node $n_j$ with weight $w$. Against different backgrounds, the meanings of these edges will be different. For example, in recommender systems, there could be a user social network which expresses the trust relations between users; in the text mining area, there could be a paper citation network which expresses the citation relations between papers.

\begin{definition}[Horizontal Network Factorization] Horizontal network factorization projects the nodes to a lower dimension space and, at the same time, keeps the original horizontal network structure as unchanged as possible. Based on the NMF, it can be expressed as
\begin{equation}
\begin{aligned}
J \left ( A \right ) =& \frac{1}{2} {\| H - AA^T \|}_F^2,
\end{aligned}
\label{nnmfh}
\end{equation}
where $A_{n \times k}$ is a nonnegative matrix with each row corresponding to a node in the network $H_{n \times n}$. We can see that the nodes in $H_{n \times n}$ are all projected to a $k$-dimensional space as $A$. The cost function in Eq. (\ref{nnmfh}) tries as far as possible to retain the network structure of $H$.
\end{definition}

\subsection{Vertical-network Factorization}

Given a vertical network $V_{n \times p}$, the element $v_{n_i, p_i}$ denotes that there is an edge between node $n_i$ and node $p_i$. Against different backgrounds, the meanings of these edges will also be different. For example, in recommender systems, there could be user-movie relations due to user ratings on movies; in the text mining area, there could be author-paper relations as a result of authors writing papers.

\begin{definition}[Vertical Network Factorization] The vertical network factorization also projects the nodes to a lower dimension space and, at the same time, keeps the original vertical network structure as unchanged as possible. Based on the NMF, it can be expressed as
\begin{equation}
\begin{aligned}
J \left ( A, X \right ) =& \frac{1}{2} {\| V - AX \|}_F^2,
\end{aligned}
\label{nnmfv}
\end{equation}
where $A_{n \times k}$ is a nonnegative matrix with rows corresponding to one kind of node in the network $V_{n \times p}$ and $X_{k \times p}$ corresponding to the other kind of node in $V_{n \times p}$. We can see that nodes with different characters in $V_{n \times p}$ are all projected to a $k$-dimensional space. The cost function in Eq. (\ref{nnmfv}) tries as far as possible to retain the network structure of $V$.
\end{definition}

Note that the nonnegativity condition of $A$ and $X$ is necessary, because the underlying components $A$ and $X$ have their own physical interpretations. In the recommender system, $A$ denotes the users' interests and $X$ denotes the movies' properties.

\subsection{Multi-level Network Factorization}

The friend relations between users will apparently impact on users' ratings on movies, and the citation relations between papers will impact on the keyword usage of each paper. Thus, we need a way to combine them.

\begin{definition}[Multi-level Network Factorization] Multi-level network factorization is the combination of Horizontal-network factorization and Vertical-network factorization, and the final latent $k$-dimensional space is shared by two factorizations. Based on the NMF, it can be expressed as
\begin{equation}
\begin{aligned}
J \left ( A, X \right ) =& \frac{1}{2} {\| V - AX \|}_F^2 + \alpha \cdot \frac{1}{2} {\| H - AA^T \|}_F^2,
\end{aligned}
\label{nnmfm}
\end{equation}
where $\alpha$ is the parameter used to adjust the weights of two parts in the cost function.
\end{definition}

It should be noted that $H$ only represents one horizontal network. However, multi-level factorization can easily be achieved by adding more items on the right side of Eq. (\ref{nnmfm}). Here, we only discuss one level horizontal network for the sake of simplicity, but a comprehensive analysis of the situations with a different number of levels will be tested in the experiment section.

One problem when conducting multi-level network factorization is how to preserve the network structures, such as community structure, degree distribution and max spanning tree, after the projection to the new and relatively small $k$-dimensional latent space. Note that the minimization of Eq. (\ref{nnmfm}) cannot ensure the different structures will definitely be preserved, as will be demonstrated in the experiments in Section \Rmnum{5}.


\section{Preserving Complex Network Structures during Multi-level Network Factorization}

In this section, we will introduce four algorithms for preserving four different network structures during multi-level network factorization. There are two levels of nodes with vertical network, $V_{n \times p}$, but only one level horizontal network, $H_{n \times n}$, will be considered for brevity.

\subsection{NNMF: Preserve the Whole Network Structure}

In this situation, all edges in the network have the same status. The cost function is the same as Eq. (\ref{nnmfm}). Since Eq. (\ref{nnmfm}) contains second-order of matrix $A$, an approximation with lower computation complexity can be made by
\begin{equation}
\begin{aligned}
 P  = {\arg\min}_{\substack{P \ge 0}}{ \Bigg \{\frac{1}{2} {\| H - PP^T \|}_F^2 \Bigg \}}
\end{aligned}
\label{nnmf0}
\end{equation}
and
\begin{equation}
\begin{aligned}
J \left ( A, X \right ) = \frac{1}{2} {\| V - AX \|}_F^2 + \alpha \frac{1}{2} {\| P - A \|}_F^2.
\end{aligned}
\label{nnmf1}
\end{equation}

The first equation is a symmetric NMF for network $H$. We can then obtain an optimized latent space which preserves the whole network structure and the new representation, $P$, of nodes by this latent space.

The derivatives of Eq. (\ref{nnmf1}) with respect to $A$ is
\begin{equation}
\begin{aligned}
\frac{\partial J}{\partial A} &= \big (  - VX^T +AXX^T \big ) + \alpha \big (  A - P \big ) \\
&=   A(XX^T + \alpha \cdot I) - (VX^T + \alpha P)
\end{aligned}
\label{nnmf1da}
\end{equation}
and according to KKT condition
\begin{equation}
\begin{aligned}
\big [ &\big (  - VX^T +AXX^T \big ) + \alpha \big (  A - P \big ) \big ]_{ij} A_{ij} = 0,
\end{aligned}
\label{nnmf1ua}
\end{equation}
the update equation is set as
\begin{equation}
\begin{aligned}
A^{t+1}_{ij} \leftarrow A^t_{ij} \sqrt{  \frac{\big [
VX^T+ \alpha P^+
\big ]_{ij} }{\big [
AXX^T+ \alpha A + \alpha P^-
\big ]_{ij} }  }
\end{aligned}
\label{nnmf1ua}
\end{equation}
where $P^+ = (|P_{ij}|+P_{ij})/2$, $P^- = (|P_{ij}|-P_{ij})/2$ and $P = P^+ - P^-$. For $X$, the derivative is
\begin{equation}
\begin{aligned}
\frac{\partial J}{\partial X} &= -A^TV + A^TAX.
\end{aligned}
\label{nnmf1dx}
\end{equation}
According to KKT condition
\begin{equation}
\begin{aligned}
\big [ -A^TV + A^TAX \big ]_{ij} X_{ij} = 0,
\end{aligned}
\end{equation}
the update equation for $X$ is
\begin{equation}
\begin{aligned}
X_{ij} \leftarrow X_{ij} \Bigg [ \frac{\big [ A^TV \big ]_{ij} }{\big [ A^TAX \big ]_{ij} } \Bigg ]^{1/2}.
\end{aligned}
\label{nnmf1ux}
\end{equation}

Since the update equation of $X$ is the same as traditional NMF, we only give the convergence proof of $A$.

\begin{proof}
According to Eq. (\ref{nnmf1ua}), give the objective function with respect to $A$ as
\begin{equation}
\begin{aligned}
F(A) =  tr \Big(& -  VX^TA^T + \frac{1}{2}AXX^TA^T + \frac{1}{2}\alpha \cdot AA^T \\
&- \alpha \cdot P^+A^T + \alpha \cdot P^-A^T \Big)
\end{aligned}
\end{equation}
and define
\begin{equation}
\begin{aligned}
G(A, A^t) &= \sum_{ij} - (VX^T)_{ij}A^t_{ij}(1+\log \frac{A_{ij}}{A^t_{ij}}) \\
			&+ \sum_{ij} \frac{(A^tXX^T)_{ij}A^2_{ij}}{2A^t_{ij}}
			 + \sum_{ij} \alpha \frac{(A^t_{ij})A^2_{ij}}{2A^t_{ij}} \\
			&- \sum_{ij} \alpha (P^+)_{ij}A^t_{ij}(1+\log \frac{A_{ij}}{A_{ij}^t}) \\
            &+ \sum_{ij} \alpha (P^-)_{ij} \frac{A_{ij}^2 + (A^t_{ij})^2}{A^t_{ij}}.
\end{aligned}
\end{equation}
Then, $G(A, A^t)$ is the auxiliary function of $F(A)$, because the conditions
\begin{equation}
\begin{aligned}
F(A^t) &= G(A^t,A^t) \\
G(A^t,A^t) &\ge G(A^{t+1},A^t) \\
G(A^{t+1},A^t) &\ge F(A^{t+1})
\end{aligned}
\label{af}
\end{equation}
are satisfied when $A^{t+1}$ takes the minimum value of $G(A,A^t)$ with respect to $A_{ij}$. The derivative of $G(A, A^t)$ is
\begin{equation}
\begin{aligned}
\frac{\partial G(A, A^t)}{\partial A_{ij}} =& - \frac{(VX^T)_{ij}A^t_{ij}}{A_{ij}}
								+ \frac{(A^tXX^T)_{ij}A_{ij}}{A^t_{ij}} \\
								&+ \alpha \frac{A^t_{ij}A_{ij}}{A^t_{ij}}
								+ \alpha  \frac{(P^-_{ij})A_{ij}}{A^t_{ij}}
                                - \alpha  \frac{P^+_{ij}A^t_{ij}}{A_{ij}}\\
\end{aligned}
\end{equation}
and set $A^{t+1}$ as the minimal value of $G(A, A^t)$ through $\frac{\partial G(A, A^t)}{\partial A_{ij}} = 0$. We have
\begin{equation}
\begin{aligned}
A^{t+1}_{ij} &= A^t_{ij} \sqrt{  \frac{\big [
VX^T+ \alpha P^+
\big ]_{ij} }{\big [
AXX^T+ \alpha A + \alpha P^-
\big ]_{ij} }  }
\end{aligned}
\label{updateA}
\end{equation}

Therefore, we know that the update of $A$ according to Eq. (\ref{nnmf1ua}) will lead to the non-increasing of $J(A,X)$.

\end{proof}

We can see from the update equations for $A$ and $X$ that, since we only consider one level network, the $A$ is influenced by the horizontal network structure in Eq. (\ref{nnmf1ua}). The update equation of $X$ is the same as Eq. (\ref{nnmf1ux}), since the new term does not impact on the derivative of cost function with respect to $X$.

Eq. (\ref{updateA}) is equal to update $A$ through the gradient descent method with a special step size \cite{4359171,NIPS2000_1861}. The corresponding step size of Eq. (\ref{updateA}) is
\begin{equation}
\begin{aligned}
&\eta_A = A_{ij} \cdot
\\
&\Bigg [\left ((AXX^T+ \alpha A + \alpha P^-)_{ij}^{1/2} + (VX^T + \alpha P^+)_{ij}^{1/2} \right )
\\
&~~~~~~\cdot \left (AXX^T+ \alpha A + \alpha P^- \right )_{ij}^{1/2} \Bigg ]^{-1}
\end{aligned}
\end{equation}
As pointed out in \cite{4359171,Lin:2007}, Eq. (\ref{updateA}) might not converge to a stationary point due to improper step size. To ensure the convergence of the update, we revise this step size \cite{4359171} as,
\begin{equation}
\begin{aligned}
&\overline{\eta}_A = \overline{A_{ij}} \cdot
\\
&\Bigg [\left ((\overline{A}XX^T+ \alpha \overline{A} + \alpha P^-)_{ij}^{1/2} + (VX^T + \alpha P^+)_{ij}^{1/2} \right )
\\
&~~~~~~\cdot \left (\overline{A}XX^T+ \alpha \overline{A} + \alpha P^- \right )_{ij}^{1/2} + \delta \Bigg ]^{-1}
\end{aligned}
\label{etaa}
\end{equation}
where
\begin{equation}
\label{maxA}
  \overline{A_{ij}} = \left\{
   \begin{array}{cl}
   A_{ij},  &if \frac{\partial J}{\partial A_{ij}} \geq 0\\
   max(A_{ij}, \sigma),  &if \frac{\partial J}{\partial A_{ij}} < 0 \\
   \end{array}
  \right.
\end{equation}
and $\delta$ and $\sigma$ are two small positive numbers.

For $X$, the corresponding step size of Eq. (\ref{nnmf1ux}) is
\begin{equation}
\begin{aligned}
&\overline{\eta}_X = \overline{X}_{ij} \cdot
\\
&\Bigg [\left ((A^TA\overline{X})_{ij}^{1/2} + (A^TV)_{ij}^{1/2} \right )
\cdot \left (A^TA\overline{X} \right )_{ij}^{1/2} + \delta \Bigg ]^{-1}
\end{aligned}
\label{etax}
\end{equation}

The whole procedure is summarized in Algorithm \ref{algnnmf}.

\begin{algorithm}[t]
\caption{NNMF}
\label{algnnmf}
\begin{algorithmic}[1]
    \REQUIRE $H$ and $V$, Maximum Iteration number: $I_{\max}$
    \ENSURE $A$ and $X$
    \STATE $P  = {\arg\min}_{\substack{P \ge 0}}{ \Bigg \{\frac{1}{2} {\| H - PP^T \|}_F^2 \Bigg \}}$;
    \WHILE{$i < I_{\max}$}
        \IF{ $\frac{\partial J}{\partial A_{ij}} < 0$ in Eq. (\ref{nnmf1da}) }
        \STATE $\overline{A_{ij}} = max(A_{ij}, \sigma)$;
        \ENDIF
        \STATE compute $\overline{\eta}_A$ by Eq. (\ref{etaa});
        \STATE $A_{ij} \leftarrow A_{ij} - \overline{\eta}_A \frac{\partial J}{\partial A_{ij}}$;
        \IF{ $\frac{\partial J}{\partial X_{ij}} < 0$ in Eq. (\ref{nnmf1dx})}
        \STATE $\overline{X_{ij}} = max(X_{ij}, \sigma)$;
        \ENDIF
        \STATE compute $\overline{\eta}_X$ by Eq. (\ref{etax});
        \STATE $X_{ij} \leftarrow X_{ij} - \overline{\eta}_X \frac{\partial J}{\partial X_{ij}}$;
        \STATE $i= i + 1$;
    \ENDWHILE
\end{algorithmic}
\end{algorithm}


\subsection{CNMF: Preserve Community Structure}

Community structure \cite{estrada2011structure,Boccaletti2006175,cmdmkd} is an important structural property of a complex network. Like the whole network structure, our idea is to project the nodes into a latent space which can preserve community structure rather than the whole network structure. Then, the problem becomes how to find a space to preserve the community structure. The Laplacian matrix, which is broadly used in spectral analysis\cite{von2007tutorial}, of the original network matrix, $H$, is considered here. One of its definition is,
\begin{equation}
\begin{aligned}
L &= D - H \\
(P, \lambda) &= svd(L)\\
\end{aligned}
\label{nnmfcL}
\end{equation}
where $D$ is the degree matrix defined as $d_{i,i} = \sum_j h_{i,j}$ and $d_{i,j} = 0 (i\ne j)$, $svd(\cdot)$ is the singular value decomposition operation, $P$ is the eigenvectors and $\lambda$ is eigenvalues.

With this latent space, $P$, the cost function can be easily designed as
\begin{equation}
\begin{aligned}
J \left ( A, X \right ) =& \frac{1}{2} {\| V - AX \|}_F^2 + \alpha \cdot \frac{1}{2} {\| P_{k} - A \|}_F^2,
\end{aligned}
\label{nnmfc}
\end{equation}
where $P_k$ is the first $k$ eigenvectors in $P$. Then, the update equations are the same as Eq. (\ref{nnmf1ua}) and Eq. (\ref{nnmf1ux}). Next, we try to prove the ability to preserve the community structure of the designed cost function.

\begin{proof}
$P$'s ability to preserve the community structure originates from graph cut theory. In this theory, separating a group of nodes into $k$ subgroups is equal to optimizing the following cost function
\begin{equation}
\begin{aligned}
&\min_{G}  RatioCut(G_1, G_2, ..., G_k)  \\
=& \min_{G} \Big \{ \frac{1}{2} \sum_{i=1}^k \frac{W(G_i, \overline{G_i})}{|G_i|}  \Big \} \\
=& \min_{Q} Tr(Q^TLQ), ~s.t.~ Q^TQ = I \\
\end{aligned}
\label{nnmfcgraphcut}
\end{equation}
where $G = {G_1, G_2, ..., G_k}$ is the partition of nodes, $|G_i|$ is the number of nodes in $G_i$, and $W$ is a designed indicator matrix (more details can be found in \cite{von2007tutorial}). With a small relaxation, the solution $Q$ in Eq. (\ref{nnmfcgraphcut}) is just the matrix that contains the first $k$ eigenvectors of $L$. That means our $P_k$ can optimize the cost function in Eq. (\ref{nnmfcgraphcut}), and then can give the best partition of the network. Therefore, the cost function in Eq. (\ref{nnmfc}) is able to preserve the community structure of the original network $H$.
\end{proof}

It should be noted that the cost function in Eq. (\ref{nnmfc}) has the same trend with
\begin{equation}
\begin{aligned}
J \left ( A, X \right ) =& \frac{1}{2} {\| V - AX \|}_F^2
+ \alpha \cdot tr(A^TLA).
\end{aligned}
\label{nnmfc2}
\end{equation}
This equation directly combines the NMF and Graph-Cut, which are commonly adopted by other researches \cite{5674058,gu2010} for graph-embedding or graph-regularization. The difference between Eq. (\ref{nnmfc}) and Eq. (\ref{nnmfc2}) is the same as the difference between Eq. (\ref{nnmf1}) and Eq. (\ref{nnmfm}).

The whole procedure is summarized in Algorithm \ref{algcnmf}.

\begin{algorithm}[t]
\caption{CNMF}
\label{algcnmf}
\begin{algorithmic}[1]
    \REQUIRE $H$ and $V$, Maximum Iteration number: $I_{\max}$
    \ENSURE $A$ and $X$
    \STATE $L = D - H$;
    \STATE $P = svd(L)$;
    \WHILE{$i < I_{\max}$}
        \IF{ $\frac{\partial J}{\partial A_{ij}} < 0$ in Eq. (\ref{nnmf1da}) }
        \STATE $\overline{A_{ij}} = max(A_{ij}, \sigma)$;
        \ENDIF
        \STATE compute $\overline{\eta}_A$ by Eq. (\ref{etaa});
        \STATE $A_{ij} \leftarrow A_{ij} - \overline{\eta}_A \frac{\partial J}{\partial A_{ij}}$;
        \IF{ $\frac{\partial J}{\partial X_{ij}} < 0$ in Eq. (\ref{nnmf1dx})}
        \STATE $\overline{X_{ij}} = max(X_{ij}, \sigma)$;
        \ENDIF
        \STATE compute $\overline{\eta}_X$ by Eq. (\ref{etax});
        \STATE $X_{ij} \leftarrow X_{ij} - \overline{\eta}_X \frac{\partial J}{\partial X_{ij}}$;
        \STATE $i= i + 1$;
    \ENDWHILE
\end{algorithmic}
\end{algorithm}


\subsection{DNMF: Preserve Degree Distribution Structure}

The degree distribution \cite{estrada2011structure,Bara15101999,Adamic24032000} is another very important structure. A node, $i$, in a network will have a number of neighbors as, $d_i$. The degree distribution can be described by a function of $n_d \sim \pi(d)$, where $n_d$ denotes the number of nodes with the same degree $d$ and $\pi(\cdot)$ is the distribution of the number of degrees. Our idea to preserve the degree distribution is to maximize the correlation between the node degree sequences of the original matrix and the new generated matrix (formed by $k$-dimensional latent space).

Suppose the network matrix $H$ is a binary matrix,
\begin{equation}
\begin{aligned}
H_{n \times n}\mathbf{1}^{n \times 1}
\end{aligned}
\end{equation}
will be a vector containing degrees of all nodes. Similarly,
\begin{equation}
\begin{aligned}
AA^T\mathbf{1}^{n \times 1}
\end{aligned}
\end{equation}
is also a vector containing degrees of all nodes in the $k$-dimensional latent space. The distance between two degree vectors can be evaluated by,
\begin{equation}
\begin{aligned}
\| H_{n \times n}\mathbf{1}^{n \times 1} -
AA^T\mathbf{1}^{n \times 1} \|_F^2
\end{aligned}
\end{equation}
If we can minimize this distance, the degree distributions in original space and new latent space will be similar. If the $H$ is not a binary matrix but a real-valued matrix, the degree sequence an be seen as a \textit{weighted degree distribution}. Then, a new cost function with this term is
\begin{equation}
\begin{aligned}
J(A,X) &= \frac{1}{2} {\| V - AX \|}_F^2 \\
&+ \frac{1}{2} \alpha \cdot \| H_{n \times n}\mathbf{1}^{n \times 1}
- AA^T\mathbf{1}^{n \times 1} \|_F^2.
\end{aligned}
\label{costfundnmf}
\end{equation}

The derivative of $A$ is
\begin{equation}\label{deriveAdnmf}
\begin{aligned}
\frac{\partial J(A,X)}{\partial A}
&=   - VX^T +AXX^T  \\
&- \alpha \cdot H\mathbf{1}^{n \times 1}\mathbf{1}^{1 \times n}A \\
&+ 2 \alpha \cdot AA^T\mathbf{1}^{n \times 1}\mathbf{1}^{1 \times n}A \\
\end{aligned}
\end{equation}

The update equation of $A$ is set
\begin{equation}
\begin{aligned}
&A_{ij} \leftarrow   A_{ij}
\\
&\cdot \Bigg [
\left ((AXX^T)^2_{ij} + 8\alpha (AA^T\mathbf{1}A)_{ij} (VX^T + \alpha H\mathbf{1}A)_{ij} \right )^{1/2}
\\
& - (AXX^T)_{ij} \Bigg ]^{\frac{1}{2}}
 / \Bigg [(AA^T\mathbf{1}^{n \times n}A)_{ij} \Bigg ]^{\frac{1}{2}}
\end{aligned}
\label{nmfdua}
\end{equation}

\begin{proof}According to Eq. (\ref{costfundnmf}), give the objective function with respect to $A$ as
\begin{equation}
\begin{aligned}
F(A) =  tr \Big( &-  YX^TA^T
\\
&+ \frac{1}{2}AXX^TA^T \\
&- \alpha \cdot H\mathbf{1}^{n \times n}AA^T \\
&+ \frac{1}{2}\alpha \cdot AA^T\mathbf{1}^{n \times n} AA^T
\Big)
\end{aligned}
\end{equation}
and define
\begin{equation}
\begin{aligned}
&~~~G(A, A^t) \\
&=  - \sum_{ij} \Bigg ( (VX^T)_{ij}A^t_{ij}(1+\log \frac{A_{ij}}{A^t_{ij}}) \Bigg ) \\
&+ \frac{1}{2}\sum_{ij} \Bigg ( \frac{(A^tXX^T)_{ij}A^2_{ij}}{A^t_{ij}} \Bigg )
\\
&- \sum_{ij} \Bigg ( \alpha \cdot (H\mathbf{1}^{n \times n}A^t)_{ij}
A^t_{ij}(1+\log \frac{A_{ij}}{A^t_{ij}}) \Bigg )
\\
&+ \frac{1}{2}\sum_{ij} \Bigg ( \alpha \cdot \frac{(A^t(A^t)^T\mathbf{1}^{n \times n}A^t)_{ij} A^4_{ij}}{(A^t_{ij})^3} \Bigg )
\\
\end{aligned}
\end{equation}
Then, $G(A, A^t)$ is the auxiliary function of $F(A)$, because Eq. (\ref{af}) is satisfied. The derivative of $G(A, A^t)$ with respect to $A$ is
\begin{equation}
\begin{aligned}
\frac{\partial G(A, A^t)}{\partial A_{ij}}
= -& \frac{(VX^T)_{ij}A^t_{ij}}{A_{ij}} \\
+& \frac{(A^tXX^T)_{ij}A_{ij}}{A_{ij}^t}
\\
-& \alpha \cdot \frac{((H\mathbf{1}^{n \times n})
A^t)_{ij}A^t_{ij}}{A_{ij}}
\\
+& 2 \alpha \cdot \frac{(AA^T\mathbf{1}^{n \times 1}\mathbf{1}^{1 \times n}A)_{ij} A^3_{ij}}{(A^t_{ij})^3}
\\				
\end{aligned}
\end{equation}
and set $A^{t+1}$ through $\frac{\partial G(A, A^t)}{\partial A_{ij}} = 0$,
and then the Eq. (\ref{nmfdua}) is derived.
\end{proof}

The revised step size for the update in Eq. (\ref{nmfdua}) is,
\begin{equation}\label{etaadnmf}
\begin{aligned}
&\overline{\eta}_A = \overline{A}_{ij} \cdot
 \Bigg [(4 \alpha \cdot (\overline{A}\overline{A}^T\mathbf{1}\overline{A})_{ij})^{1/2} \\
& - \Bigg  ((\overline{A}XX^T)^2_{ij} + 8\alpha (\overline{A}\overline{A}^T\mathbf{1}\overline{A})_{ij} \cdot (VX^T + \alpha \cdot H\mathbf{1}\overline{A})_{ij} \Bigg )^{1/2}
\\
&- (\overline{A}XX^T)_{ij} \Bigg ]
/ \Bigg [(4 \alpha \cdot (\overline{A}\overline{A}^T\mathbf{1}\overline{A})_{ij})^{1/2}
\\
&\cdot (- VX^T +\overline{A}XX^T
- \alpha \cdot H\mathbf{1}\overline{A}
+ 2 \alpha \cdot \overline{A}\overline{A}^T\mathbf{1}\overline{A}) + \delta\Bigg ]
\end{aligned}
\end{equation}
where $\overline{A}$ satisfies Eq. (\ref{maxA}) with $\frac{\partial J}{\partial A}$ replaced by Eq. (\ref{deriveAdnmf}).

The whole procedure is summarized in Algorithm \ref{algdnmf}.

\begin{algorithm}[t]
\caption{DNMF}
\label{algdnmf}
\begin{algorithmic}[1]
    \REQUIRE $H$ and $V$, Maximum Iteration number: $I_{\max}$
    \ENSURE $A$ and $X$
    \WHILE{$i < I_{\max}$}
        \IF{ $\frac{\partial J}{\partial A_{ij}} < 0$ in Eq. (\ref{deriveAdnmf}) }
        \STATE $\overline{A_{ij}} = max(A_{ij}, \sigma)$;
        \ENDIF
        \STATE compute $\overline{\eta}_A$ by Eq. (\ref{etaadnmf});
        \STATE $A_{ij} \leftarrow A_{ij} - \overline{\eta}_A \frac{\partial J}{\partial A_{ij}}$;
        \IF{ $\frac{\partial J}{\partial X_{ij}} < 0$ in Eq. (\ref{nnmf1dx})}
        \STATE $\overline{X_{ij}} = max(X_{ij}, \sigma)$;
        \ENDIF
        \STATE compute $\overline{\eta}_X$ by Eq. (\ref{etax});
        \STATE $X_{ij} \leftarrow X_{ij} - \overline{\eta}_X \frac{\partial J}{\partial X_{ij}}$;
        \STATE $i= i + 1$;
    \ENDWHILE
\end{algorithmic}
\end{algorithm}


\subsection{TNMF: Preserve Max Spanning Tree Structure}

The max spanning tree \cite{estrada2011structure,Boccaletti2006175} of network $H$ is first mined,
\begin{equation}
\begin{aligned}
T \circledast H = T^H
\end{aligned}
\label{mst}
\end{equation}
where $\circledast$ denotes the Hadamard (element-wise) product and $T^H$ is the mined max spanning tree of network $H$ and $T$ is called the \textit{tree-mask matrix} which is a binary matrix with $t_{i,j}=1$ if it is in $T^H$ otherwise $t_{i,j}=0$, and $\overline{T}$ is the complement of $T$ with $t_{i,j}=1$ if it is not in $T^H$ otherwise $t_{i,j}=0$.

The cost function is
\begin{equation}
\begin{aligned}
J(A,X) &= \frac{1}{2} {\| V - AX \|}_F^2 \\
&~~~- \alpha \cdot \frac{1}{4} {\| T \circledast (AA^T) - \overline{T} \circledast (AA^T)) \|}^2_F.
\end{aligned}
\label{costtnmf}
\end{equation}

The derivative with respect to $A$ is
\begin{equation}\label{deriatnmf}
\begin{aligned}
\frac{\partial J(A,X)}{\partial A} &= \big (  - VX^T +AXX^T \big ) \\
  &+ \alpha \cdot  (\overline{T} - T) \circledast (AA^T)A
\end{aligned}
\end{equation}

The update equation for $A$ is
\begin{equation}
\begin{aligned}
&A_{ij} \leftarrow   A_{ij} \cdot
\Bigg [
\Big ((AXX^T)^2_{ij}
\\
&+ 4\alpha (\overline{T}\circledast (AA^T)A)_{ij} (VX^T + \alpha T \circledast (AA^T)A)_{ij} \Big )^{1/2}
\\
& - (AXX^T)_{ij} \Bigg ]^{\frac{1}{2}}
 / \Bigg [2\alpha(\overline{T} \circledast AA^T A) \Bigg ]^{\frac{1}{2}}
\end{aligned}
\label{tnmfua}
\end{equation}
The convergence proof of this update equation is given as follows.

\begin{proof}
According to Eq. (\ref{costtnmf}), give the objective function with respect to $A$
\begin{equation}
\begin{aligned}
F(A) &=  tr \Big( -  VX^TA^T +\frac{1}{2}AXX^TA^T \\
&~~~~+ \frac{1}{4} \alpha \cdot \big ( (\overline{T} - T) \circledast AA^TAA^T \Big)
\end{aligned}
\end{equation}
and define
\begin{equation}
\begin{aligned}
&G(A, A^t) = -  \sum_{ij} \left ( (VX^T)_{ij}A^t_{ij}(1+\log \frac{A_{ij}}{A^t_{ij}}) \right ) \\
&+ \frac{1}{2}\sum_{ij} \left ( \frac{(A^tXX^T)_{ij}A^2_{ij}}{(A^t_{ij})} \right ) \\
& -\frac{1}{4}\alpha \sum_{ijkl} \left ( T_{il}A^t_{il}A^t_{kl}A^t_{ik}A^t_{ij} \left (1 + \log\frac{A_{il}A_{kl}A_{ik}A_{ij}}{A^t_{il}A^t_{kl}A^t_{ik}A^t_{ij}} \right ) \right )
\\
& +\frac{1}{4}\alpha \sum_{ij} \left ( \frac{(\overline{T} \circledast (A^t(A^t)^T)A^t)_{ij} (A_{ij})^4}{(A^t_{ij})^3} \right ).
\end{aligned}
\end{equation}

Then, $G(A, A^t)$ is the auxiliary function of $F(A)$, because Eq. (\ref{af}) is satisfied. The derivative of $G(A, A^t)$ with respect to $A$ is
\begin{equation}
\begin{aligned}
\frac{\partial G(A, A^t)}{\partial A_{ij}} =
&-  \frac{(VX^T)_{ij}A^t_{ij}}{A_{ij}} \\
&+ \frac{(A^tXX^T)_{ij}A_{ij}}{(A^t_{ij})}  \\
& -\alpha \frac{(T \circledast (A^t(A^t)^T)A^t)_{ij} A^t_{ij}}{A_{ij}}   \\
& +\alpha \frac{(\overline{T} \circledast (A^t(A^t)^T)A^t)_{ij} A_{ij}^3}{(A^t_{ij})^3}
\end{aligned}
\end{equation}
and set $A^{t+1}_{ij}$ as the minimal value of $G(A, A^t)$ through $\frac{\partial G(A, A^t)}{\partial A_{ij}} = 0$. Then, we get Eq. (\ref{tnmfua}).
Therefore, we know that the update of $A$ according to Eq. (\ref{tnmfua}) will lead to the non-increasing of $J(A,X)$.

\end{proof}

The revised step size for the update in Eq. (\ref{nmfdua}) is,
\begin{equation}\label{etaatnmf}
\begin{aligned}
&\overline{\eta}_A = \overline{A}_{ij} \cdot
 \Bigg [\left (2\alpha \cdot (\overline{T}\circledast (\overline{A}\overline{A}^T)\overline{A} \right )_{ij}^{1/2} - \Bigg  ((\overline{A}XX^T)^2_{ij}
 \\
 &+ 4\alpha (\overline{T}\circledast (\overline{A}\overline{A}^T)\overline{A})_{ij}  (VX^T + \alpha T\circledast (\overline{A}\overline{A}^T)\overline{A})_{ij} \Bigg )^{1/2}
\\
&- (\overline{A}XX^T)_{ij} \Bigg ]
/ \Bigg [(2\alpha \cdot \left (\overline{T}\circledast (\overline{A}\overline{A}^T)\overline{A} \right )_{ij}^{1/2}
\\
&\cdot \left (- VX^T +\overline{A}XX^T
+ \alpha \cdot (\overline{T} -T)\circledast (\overline{A}\overline{A}^T) \overline{A} \right )_{ij} + \delta\Bigg ]
\end{aligned}
\end{equation}
where $\overline{A}$ satisfies Eq. (\ref{maxA}) with $\frac{\partial J}{\partial A}$ replaced by Eq. (\ref{deriatnmf}).


The whole procedure is summarized in Algorithm \ref{algtnmf}.

\begin{algorithm}[t]
\caption{TNMF}
\label{algtnmf}
\begin{algorithmic}[1]
    \REQUIRE $H$ and $V$, Maximum Iteration number: $I_{\max}$
    \ENSURE $A$ and $X$
    \STATE Obtain Max spanning tree $T$ of $H$;
    \WHILE{$i < I_{\max}$}
         \IF{ $\frac{\partial J}{\partial A_{ij}} < 0$ in Eq. (\ref{deriatnmf}) }
        \STATE $\overline{A_{ij}} = max(A_{ij}, \sigma)$;
        \ENDIF
        \STATE compute $\overline{\eta}_A$ by Eq. (\ref{etaatnmf});
        \STATE $A_{ij} \leftarrow A_{ij} - \overline{\eta}_A \frac{\partial J}{\partial A_{ij}}$;
        \IF{ $\frac{\partial J}{\partial X_{ij}} < 0$ in Eq. (\ref{nnmf1dx})}
        \STATE $\overline{X_{ij}} = max(X_{ij}, \sigma)$;
        \ENDIF
        \STATE compute $\overline{\eta}_X$ by Eq. (\ref{etax});
        \STATE $X_{ij} \leftarrow X_{ij} - \overline{\eta}_X \frac{\partial J}{\partial X_{ij}}$;
        \STATE $i= i + 1$;
    \ENDWHILE
\end{algorithmic}
\end{algorithm}


\section{Experimental Results and Analysis}

This section is composed of two parts: experiments on synthetic data and experiments on real-world data. The first part is used to verify the correctness of our proposed algorithms. The second part is used to show the usefulness of our work. For the sake of brevity, abbreviations are used as follows: NNMF for the algorithm proposed in Section \Rmnum{4}(A) to preserve the whole network structure; CNMF for the algorithm proposed in Section \Rmnum{4}(B) to preserve the community structure; DNMF for the algorithm proposed in Section \Rmnum{4}(C) to preserve the degree distribution structure; TNMF for the algorithm proposed in Section \Rmnum{4}(D) to preserve the max spanning tree structure.

\subsection{Experiments on Synthetic data}

In this section, we randomly construct two matrices, $V$ and $H$, to verify the proposed algorithms, including the convergence of algorithms and the ability to preserve the desired network structures.

\subsubsection{Convergence analysis}

\begin{figure*}[!htbp]
\centerline{\includegraphics[scale=0.6]{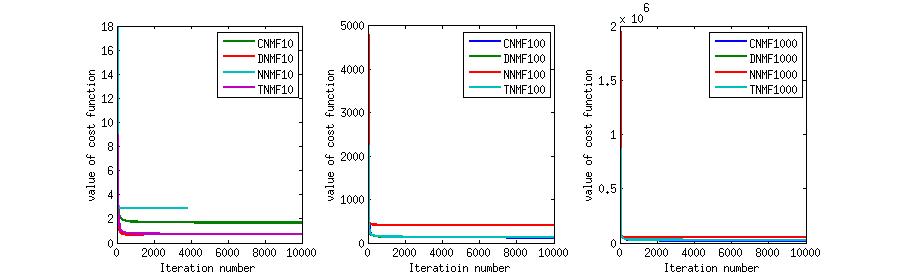}}
\caption{The values of cost functions under the designed update equations. Three figures denote the different values of the dimension numbers of latent space, 10, 100 and 1000, respectively.}
\label{fig:convergence}
\end{figure*}

Although we have given the proofs for the convergence of the proposed algorithms in Section \Rmnum{4}, the real values of cost functions are given in this section to show the convergence of the algorithms under the designed update equations. First, the matrices, $V$ and $H$, are randomly generated. Four algorithms are used to factorize them. The value after each iteration is recorded. The number of iterations is set as 10,000. It should be noted that the iterations may stop before this number is reached due to the stationary (smaller than $1e-10$) between the values of the cost function before and after one iteration. We can see from Fig. \ref{fig:convergence} that the values of the cost functions will decrease under the designed update equations of the four algorithms in line with our proofs.

\subsubsection{Test for community structure preservation}


For a given pair of matrices $<V, H>$, $H$ denotes a network. Suppose we want to factorize $<V, H>$ and preserve the community structure of network $H$ at the same time. To show the ability of CNMF to keep the community structure of the original network $H$, we compare the communities of the re-constructed networks ($H_{NNMF}$ and $H_{DNMF}$) with the communities of the original network $H$. For quantification purposes, we use clustering evaluation metrics.
The evaluation metrics of document clustering are Jaccard Coefficient (JC), Folkes\&Mallows (FM) and F1 measure (F1). Given a clustering result,
\begin{itemize}
\item $a$ is the number of two points that are in the same cluster of both benchmark and clustering results;
\item $b$ is the number of two points that are in the same cluster of benchmark but in different clusters of clustering results;
\item $c$ is the number of two points that are not in the same cluster of both benchmark but in the same cluster of clustering results.
\end{itemize}
and three metrics are computed by equations in Table \ref{tb:em} (bigger means better).

We randomly generate 1000 pairs of matrices $<V, H>$ for each size (10 and 100). The comparison between the communities of re-constructed network with the communities of the original network are shown in Table \ref{tb:com}. From these results, we can draw the conclusion that CNMF preserves the community structure better than NNMF.

\begin{table}[!thb]%
\centering
\renewcommand{\arraystretch}{3}
\caption{Evaluation metrics of document clustering/communities}{%
\begin{tabular}{c|c}
\hline
Jaccard Coefficient   &~~~~$JC=\frac{a}{a+b+c}$~~~~    \\\hline
Folkes \& Mallows    &~~~~$FM=\left( \frac{a}{a+b} \cdot \frac{a}{a+c} \right)^{1/2}$~~~~    \\\hline
F1 measure     &~~~~$F1=\frac{2a^2}{2a^2+ac+ab}$~~~~        \\\hline
\end{tabular}}
\label{tb:em}
\end{table}%

\begin{table*}[!thb]%
\centering
\renewcommand{\arraystretch}{1.3}
\caption{Comparison between NNMF and CNMF on preserving community structure}{%
\begin{tabular}{c|c|c|c|c|c|c}
\hline
\multirow{2}{*}{Algorithm} &\multicolumn{3}{c|}{n=10 (K=3)} &\multicolumn{3}{c}{n=100 (K=10)}  \\\cline{2-7}
        &JC     &FM    &F1   &JC    &FM    &F1    \\\hline
NNMF    &$0.1914 \pm 0.0642$      &$0.3240 \pm 0.0930$      &$0.3165 \pm 0.0879$  &$0.1894 \pm 0.0200$       &$0.3701 \pm 0.0487$       &$0.3180 \pm 0.0282$     \\\hline
CNMF    &$0.2542 \pm 0.0667$      &$0.5172 \pm 0.0885$      &$0.4615 \pm 0.0870$  &$0.3437 \pm 0.0952$      &$0.5209 \pm 0.0983$      &$0.5044 \pm 0.1022$    \\\hline
\end{tabular}}
\label{tb:com}
\end{table*}%

\subsubsection{Test for degree distribution preservation}

A pair of matrices are randomly generated $<V, H>$. $H$ is the original network whose degree distribution we want to keep. After running both NNMF and DNMF, we obtain $A_{NNMF}$ and $A_{DNMF}$ which can re-construct the network by $H_{NNMF} = A_{NNMF}A_{NNMF}^T$ and $H_{DNMF} = A_{DNMF}A_{DNMF}^T$. To show the ability of DNMF to retain the degree distribution of the original network $H$, we compute the correlation between the degree distributions of the re-constructed networks ($H_{NNMF}$ and $H_{DNMF}$) and the original network $H$. If the degree distribution of $H_{DNMF}$ has larger correlation coefficient with the degree distribution of $H$ than the degree distribution of $H_{NNMF}$, we can draw the conclusion that DNMF preserves degree distribution better than $H_{NNMF}$. Here, we consider two matrix sizes: 10 and 100. For each size, we randomly generate 1000 pairs of matrices. The number of hidden factors is set as $3$ and $10$. The average and standard deviation of results are given in Table \ref{tb:dd}.
These numbers show that the DNMF preserves more of the network degree distribution structure than NNMF.

\begin{table}[!thb]%
\centering
\renewcommand{\arraystretch}{1.3}
\caption{Comparison between NNMF and DNMF on preserving of degree distribution structure}{%
\begin{tabular}{c|c|c}
\hline
Algorithm   &n = 10 (K=3)  &n = 100 (K=10)  \\\hline
NNMF    &$0.0658 \pm 0.3314$   &$0.0045 \pm 0.0980$  \\\hline
DNMF    &$0.5548 \pm 0.2531$   &$0.9079 \pm 0.0190$   \\\hline
\end{tabular}}
\label{tb:dd}
\end{table}%


\subsubsection{Test for max spanning tree preservation}

\begin{figure*}[!th]
\centerline{\includegraphics[scale=0.5]{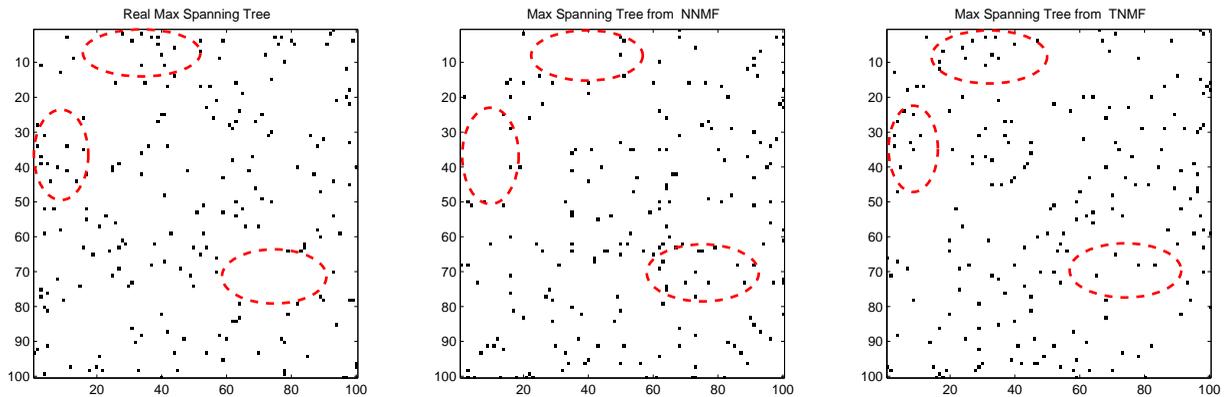}}
\caption{Comparison between NNMF and TNMF on the ability to preserve the max spanning tree. Each point in the figure denotes a link in the network. The left figure denotes the real max spanning tree of the original network; the centre figure denotes the max spanning tree from NNMF; the right figure denotes the max spanning tree from TNMF. Parts of them are highlighted as an ellipse(red). We can make an approximate comparison through these three ellipses.}
\label{fig:tree}
\end{figure*}

For the max spanning tree, we first randomly generate a pair of $V_{100 \times 100}$ and $H_{100 \times 100}$. Using NNMF and TNMF to factorize $V$ and $H$ obtain two $A_{NNMF}$ and $A_{TNMF}$ which can re-construct the network $H$ by $H_{NNMF} = A_{NNMF}A_{NNMF}^T$ and $H_{TNMF} = A_{TNMF}A_{TNMF}^T$. After extracting max spanning trees of three networks $H$, $H_{NNMF}$ and $H_{TNMF}$, we can compare the similarity between the re-constructed trees ($T_{NNMF}$ and $T_{TNMF}$) and the benchmark $T_H$. An illustrative example is shown in Fig. \ref{fig:tree}, in which each point in the matrix denotes an edge between two nodes. To quantify the ability of TNMF to retain the tree structure, we randomly generated 1000 pairs of $V$ and $H$ for each size: 10 and 100. For each pair $<V,H>$, we run both NNMF and TNMF, and compute the similarity between the re-constructed trees with the benchmark tree by the number of overlap edges as
\begin{equation}
\begin{aligned}
s(T_H, T_{NNMF}) = sum(T_H \circledast T_{NNMF})
\end{aligned}
\end{equation}
where $sum(M)$ is to count the number of $1$ in matrix $M$. The number of hidden factors are set as $3$ and $10$.
If the re-constructed max spanning tree from TNMF has more overlap edges with the original max spanning tree than the tree from NNMF, we draw the conclusion that TNMF is able to preserve the max spanning tree better than NNMF. The results are shown in Table \ref{tb:tree}. The $max$ in the table denotes the maximum number of edges in the max spanning tree. For a network with $n$ nodes, the maximum number of edges in the max spanning tree is $n-1$.

\begin{table}[!thb]%
\centering
\renewcommand{\arraystretch}{1.3}
\caption{Comparison between NNMF and TNMF on preserving max spanning tree structure}{%
\begin{tabular}{c|c|c}
\hline
Algorithm   &n = 10 (max = 9)  &n = 100 (max = 99)  \\\hline
NNMF    &$2.5 \pm 1.3$   &$21.4 \pm 6.3$   \\\hline
TNMF    &$6.2 \pm 0.7$   &$56.1 \pm 7.2$     \\\hline
\end{tabular}}
\label{tb:tree}
\end{table}%

\subsection{Experiments on Real-world Data}

The above section shows the convergence and abilities to preserve the desired network structures. In this section, we will compare the efficiency of the proposed algorithms for real-world applications, including document clustering and recommendation.

\subsubsection{Document Clustering with One-level Network}

Our document data is \emph{Cora} \footnote{http://linqs.cs.umd.edu/projects/projects/lbc/}, which is a public dataset and consists of 2708 scientific publications classified into one of seven classes. One-level network is composed by a horizontal network and a vertical network. The horizontal network, $H_{2708 \times 2708}$, is the citation relations between publications and it consists of 5429 links. This vertical network, $V_{2708 \times 1433}$, is constructed by the mapping relation between documents and words. The detailed statistics are shown in Table \ref{tb:ds1}.

\begin{table}[!thb]%
\centering
\renewcommand{\arraystretch}{1.3}
\caption{Statistics of \emph{Cora} dataset}{%
\begin{tabular}{c|c}
\hline
~~~Number of documents~~~   &~~~~2,708~~~~    \\\hline
~~~Number of keywords~~~    &~~~~1,433~~~~    \\\hline
~~~Number of links~~~    &~~~~5,429~~~~    \\\hline
~~~Number of classes~~~     &~~~~7~~~~        \\\hline
\end{tabular}}
\label{tb:ds1}
\end{table}%

After the factorization of $V_{2708 \times 1433}$ and $H_{2708 \times 2708}$ through our proposed algorithms, NNMF, CNMF, DNMF and TNMF, the documents are projected into a latent space and then given new representations. It is believed that the different classes are formed as a result of the intrinsics of the documents, so if the discovered latent space is good enough, it will cause the documents to cluster into these seven classes. According to this idea, we conduct the document clustering through the learned matrix $A$ (the new representations of documents) by the k-means clustering algorithm. To compare the efficiency, we also implement standard NMF which does not consider the horizontal network $H_{2708 \times 2708}$ and Relational Topic Model (RTM)\cite{chang2009relational} which is a successful probabilistic Bayesian model for the $H_{2708 \times 2708}$ and $V_{2708 \times 1433}$. RTM can also be seen as a method for the factorization but from a probabilistic view, which also does not consider the network structure.

\begin{figure*}[!th]
\centerline{\includegraphics[scale=0.52]{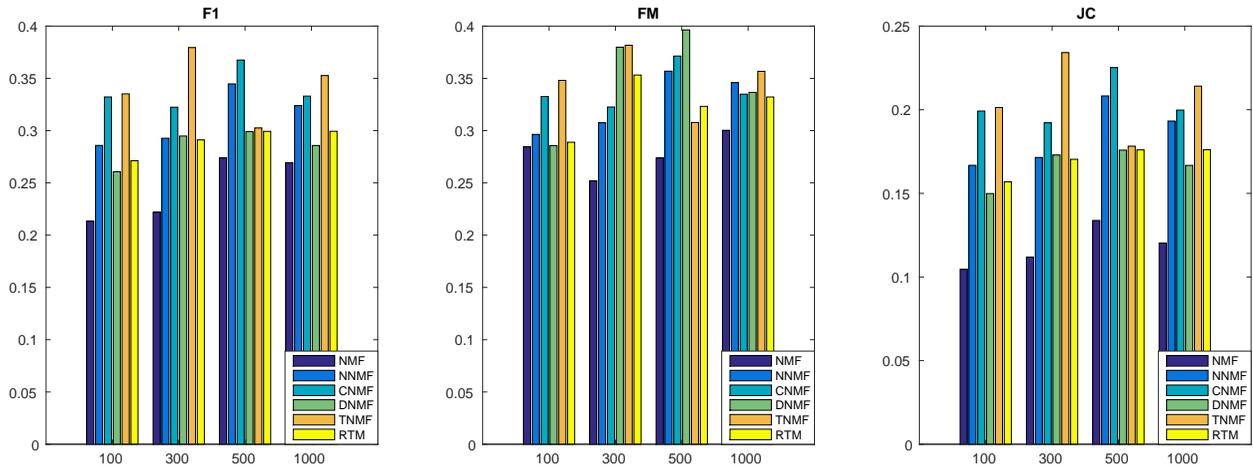}}
\caption{Comparison of document clustering on different values of $K$. The influence of $K$ values with $\alpha=0.1$.($K=100$, $K=300$, $K=500$, and $K=1000$) }
\label{fig:cora}
\end{figure*}

The final results are shown in Fig. \ref{fig:cora}. Three subfigures denote three clustering result comparisons by the metric in Table \ref{tb:em}. We have tested four numbers of factors: $K=100$, $K=300$, $K=500$ and $K=1000$. In each subfigure in Fig. \ref{fig:cora}, we have compared the results from NNMF, CNMF, DNMF, TNMF, NMF and RTM on the clustering evaluation metric. Except for NMF, the algorithms all consider the effects from the horizontal network $H_{2708 \times 2708}$. From this result, we can see that NMF achieves the worst performance compared to others. Thus, we can draw the conclusion that incorporating the citation network is helpful for the clustering of the publications. Except for DNMF, which has similar performance to RTM, NNMF, CNMF and TNMF are better than RTM on this document clustering task. Notably, CNMF and TNMF achieve the best performance of all the algorithms. The reason is that the community structure of CNMF is beneficial for the clustering because it encourages `similar' nodes to cluster together. For example, two documents $d_i$ and $d_j$ are in the same community in network $H$ due to their `similarity'. Retaining the community structure will make it more possible for $d_i$ and $d_j$ to remain within the same cluster under the new factor representations. In the TNMF, preserving max spanning tree encourages the most important relations of all the nodes/documents. These relations in the tree can be seen as the `bones' of a network, which determine the weighted distances between the nodes (documents). Therefore, the TNMF can benefit for the document clustering. It should be noted that although the network structure will influence the final clustering results, this influence will be under constraint from the document-word mapping network $V$. Each document will exhibit two natures from two networks: $H$ and $V$. If the two natures are consistent, the constraint from $V$ will help to enhance the learned network structure from $H$; If the two natures are not consistent, there will be a contradiction between $H$ and $V$, which will prevent the network structure learning from $H$.

\subsubsection{Recommendation with One-level Network}

A public dataset is adopted, \textit{Lastfm}\footnote{http://labrosa.ee.columbia.edu/millionsong/lastfm},  which is a commonly used dataset for evaluating algorithms for recommender systems. In this dataset, a vertical network $V_{1892 \times 17632}$ is formed by users and artists, and $V_{ij}$ represents the count of user $i$ listening artist $j$. The horizontal network $H_{1892 \times 1892}$ is the user friend network. The statistics are shown in Table \ref{tb:rs}. A certain number of user-artist pairs are retained as the test data. The task is to predict the counts of users listening to these artists.

\begin{table}[!thb]%
\centering
\renewcommand{\arraystretch}{1.3}
\caption{Statistics of \textit{Lastfm} dataset}{%
\begin{tabular}{c|c}
\hline
number of users   &~~~~$1892$~~~~    \\\hline
number of artists    &~~~~$17632$~~~~    \\\hline
number of friend relations     &~~~~$12717$~~~~        \\\hline
test user-artist pairs &~~~~$5000$~~~~        \\\hline
\end{tabular}}
\label{tb:rs}
\end{table}%

The evaluation metric is Mean Absolute Error (MAE) which is the simplest and the most intuitive. The definition is
\begin{equation}\label{mae}
\begin{aligned}
MAE = \frac{1}{N} \sum_{u,i} |\widehat{r}_{u,i} - r_{u,i}|,
\end{aligned}
\end{equation}
where $N$ is the number of test user-artist parts, $r_{u,i}$ is the real count and $\widehat{r}_{u,i}$ is the predicted count. The correlation coefficient is computed by
\begin{equation}\label{rho}
\begin{aligned}
\rho = \frac{\sum_i(r_i-\bar{r})(\widehat{r}_i-\bar{\widehat{r}})}{\sqrt{\sum_i (r_i-\bar{r})^2 \sum_i(\widehat{r}_i-\bar{\widehat{r}})^2}}.
\end{aligned}
\end{equation}

\begin{figure*}[!th]
\centerline{\includegraphics[scale=0.55]{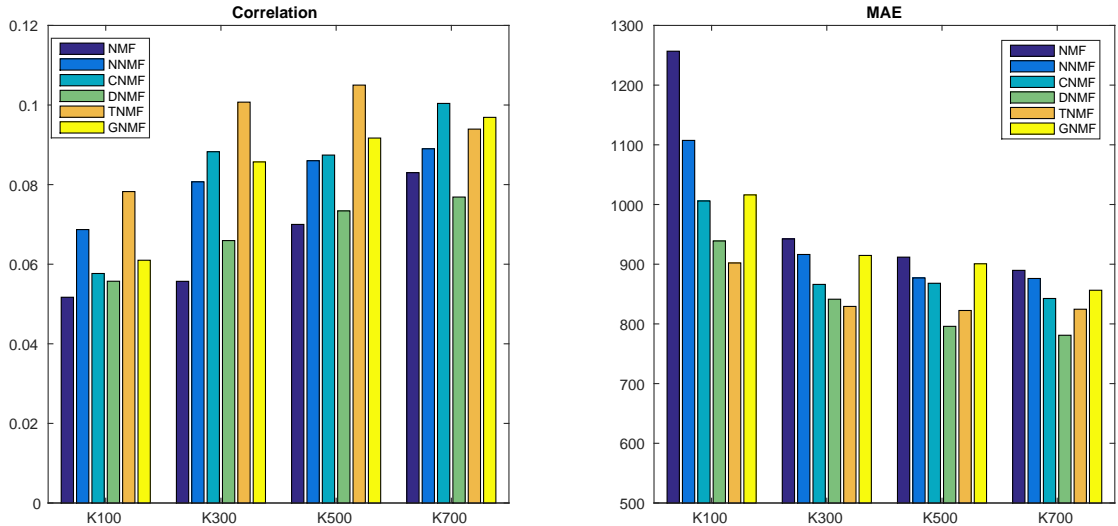}}
\caption{Comparison of recommendation on different values of $K$ ($K=100$, $K=300$, $K=500$, and $K=700$) with $\alpha=10$. }
\label{fig:lastfmk}
\end{figure*}

To show the performance of the proposed algorithms on the recommendation, we compared them with a state-of-the-art method: Graph regularized NMF (GNMF) \cite{5674058,gu2010}. The results are shown in Fig. \ref{fig:lastfmk}. The standard NMF (without the horizontal network $H_{1892 \times 1892}$) has the worst performance of all the algorithms, which is similar to the document clustering in Section V.B(1). We can therefore draw the conclusion that incorporating user friend network $H_{1892 \times 1892}$ improves the performance of the recommendation. GNMF has the same property as CNMF for preserving the community structure of $H_{1892 \times 1892}$, as discussed in Section IV.B. The results in Fig. \ref{fig:lastfmk} also show that GNMF has similar performance to CNMF. Although DNMF has good performance on MAE, the correlation with DNMF is the worst of all the proposed algorithms and the state-of-the-art GNMF.
This reflects that DNMF predicts accurate values for some test data but not all data. The reason is that DNMF preserves the network degrees of all the nodes/users. The nodes with relatively large degrees will tend to have large weights in the weighted degree distribution, and DNMF will have a bias toward these nodes during factorization. Therefore, DNMF tends to achieve good results for the nodes/users with big degrees. In all the algorithms, TNMF achieves the best performance both on MAE and Correlation. As discussed in Section V.B(1), preserving max spanning tree encourages the most important relations (the `bones' of a network) of all the nodes/users. These relations reflect not only the `distances' between nodes but also the degrees of nodes.

\subsubsection{Document Clustering with Two-level Network}

First, we introduce the dataset we use. Documents are a collection of papers from \emph{CiteSeer}\cite{sen:aimag08}. There are 3312 papers in the whole corpus. Each paper is represented by a binary vector using words. The labels of these papers are set as their research areas, such as AI (Artificial Intelligence), ML (Machine Learning), Agents, DB (Database), IR (Information Retrieval) and HCI (Human-Computer Interaction). The statistics are shown in Table \ref{tb:dsd2}.

\begin{table}[!thb]%
\centering
\renewcommand{\arraystretch}{1.3}
\caption{Statistics of \textit{CiteSeer} dataset}{%
\begin{tabular}{c|c}
\hline
~~~Number of documents~~~   &~~~~3,312~~~~    \\\hline
~~~Number of keywords~~~    &~~~~3,703~~~~    \\\hline
~~~Number of classes~~~     &~~~~6~~~~        \\\hline
\end{tabular}}
\label{tb:dsd2}
\end{table}%

A two-level network is composed by two horizontal networks $H_{3312 \times 3312}$ and $H_{3703 \times 3703}$ and one vertical network $V_{3312 \times 3703}$, and they are constructed as following. The first level horizontal network $H_{3312 \times 3312}$ is a paper citation network, which is formed by the citation relations between papers. The second level horizontal network $H_{3703 \times 3703}$ is the keyword concurrence network, which is formed by the concurrence relations between keywords. The vertical network $V_{3312 \times 3703}$ is the mapping relation between documents and keywords (there will be a link between a keyword and a document if this keyword shows in this document).

It should be noted that we keep only one type of structure for two horizontal networks. For example, CNMF only keeps the community structures of both networks, hence Eq. (\ref{nnmfc}) only adds $\alpha \cdot \frac{1}{2} \| P_X - X \|_F^2$. For brevity, the same coefficient is used for both networks.


\begin{figure*}[!th]
\centerline{\includegraphics[scale=0.65]{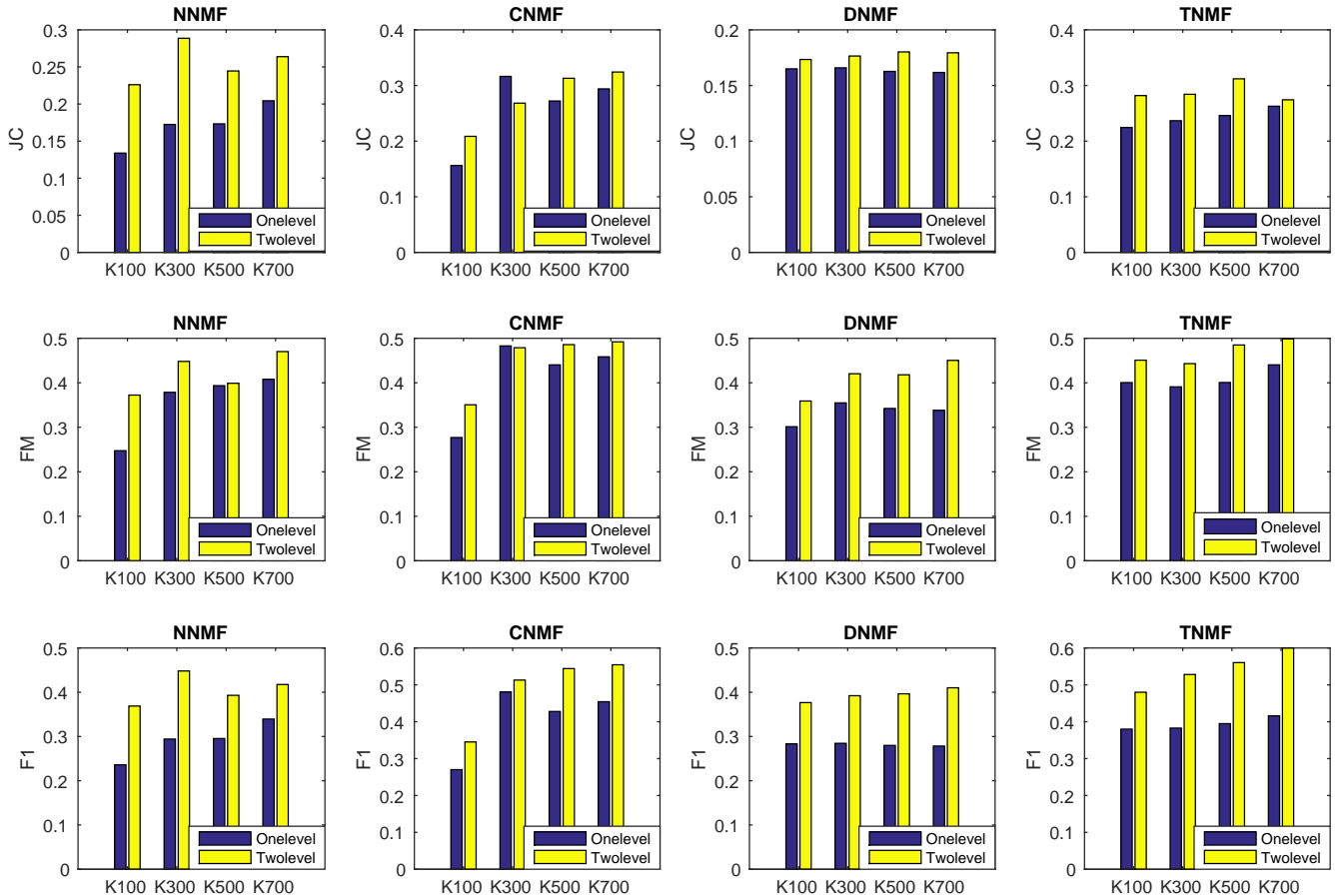}}
\caption{Comparison of the influence from one-level network (with one horizontal network and one vertical network) and two-level network (with two horizontal networks and one vertical network) on document clustering with $\alpha=0.1$ and $K=100$, $K=300$, $K=500$ and $K=700$.}
\label{fig:twolevelnetwork}
\end{figure*}

Here, we compare the performances between the one-level network $H_{3312 \times 3312}$ and the two-level network $H_{3312 \times 3312}$ and $H_{3703 \times 3703}$. The comparisons on three evaluation metrics are shown in Fig. \ref{fig:twolevelnetwork}. Except CNMF with $K=300$, two-level network outweighs the one-level network. It means that incorporating the keyword co-occurrent network can improve the document clustering task with only the document citation network.

\subsubsection{Recommendation with Two-level Network}

\begin{figure*}[!th]
\centerline{\includegraphics[scale=0.65]{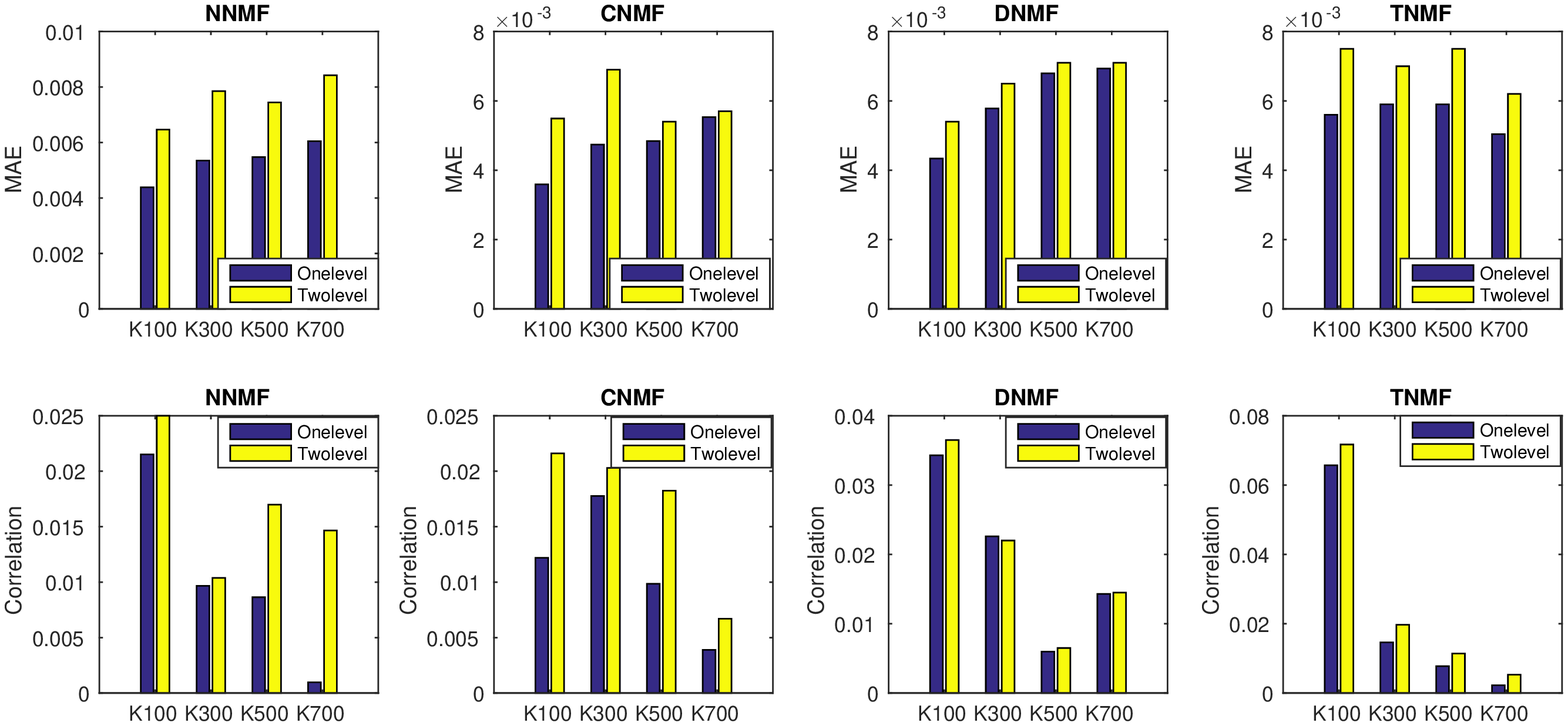}}
\caption{Comparison of the influence from one-level network (with one horizontal network and one vertical network) and two-level network (with two horizontal networks and one vertical network) on recommendation with $\alpha=10$ and $K=100$, $K=300$, $K=500$ and $K=700$.}
\label{fig:rstwolevel}
\end{figure*}

The datset used for recommendation with two-level network is \emph{Delicious}. This dataset is collected from the Delicious website\footnote{http://www.delicious.com}, which records the bookmark options of users on webpages/URLs in this website according to users' interests  \cite{Cantador:RecSys2011}. We filter this dataset by keeping the URLs that are marked by at least three users, and the number of URL is 5633. The links between URLs are generated by their tags. In the Delicious website, each URL will get tags from users, and these tags will give an hint for the content/semantics of this URL. We use these tags to compute the content similarity between URLs with tag-vector representations through cosine similarity metric. The statistics of the dataset are shown in Table. \ref{tb:rstwolevel}.
In this dataset, vertical network $V_{1867 \times 5633}$ is formed by users and URLs. The horizontal networks are: the user friend network $H_{1867 \times 1867}$ and the content similarity network $H_{5633 \times 5633}$. A certain number of user-URL pairs are kept to be the test data. The evaluation metrics are in Eq. (\ref{mae}) and Eq. (\ref{rho}). As shown in Fig. \ref{fig:rstwolevel}, the two-level network still outweighs the one-level network.

\begin{table}[!thb]%
\centering
\renewcommand{\arraystretch}{1.3}
\caption{Statistics of \textit{Delicious} dataset}{%
\begin{tabular}{c|c}
\hline
number of users   &~~~~$1867$~~~~    \\\hline
number of URLs    &~~~~$5633$~~~~    \\\hline
test user-URL pairs &~~~~$2000$~~~~        \\\hline
\end{tabular}}
\label{tb:rstwolevel}
\end{table}%

Despite the good performances on both document clustering and recommendation tasks, we still want to give an advice that it should be very careful to combine the different networks: vertical network and different horizontal networks. For example, suppose there are a user-movie vertical network and a movie network for the recommendation task, and the movie network is formed by the distance between the releasing dates of the movies. We know that the releasing date of a movie will not highly impact on the rating from users. In this situation, the combination of two networks may not improve the rating prediction due to the `noise' from the movie network. Since the information in the movie network will not help the prediction of the ratings, the constraint from movie network will decrease the learned information from the user-movie vertical network. However, if the formation of the movie network is replaced by another strategy: the movies with similar director and actors tend to have links, this new movie network will help to predict the user ratings on movies different from the old movie network. In this new situation, the combination of the movie network with user-movie network will help to improve the rating prediction. Therefore, the consistence between the different networks is important and should be considered before using the multilevel network model.


\section{Conclusions and future study}

In this paper, we have introduced a general multi-level network model, and proposed four algorithms for multi-level network factorization with four different network structure constraints. The network structural constraints have been incorporated into the cost function of traditional nonnegative matrix factorization. The optimization of the new cost function has constructed a new latent space and projected all nodes in different levels into this new latent space. At the same time, the projected nodes will, as far as possible, retain the original network structure. Four algorithms with their convergence proofs have been carefully designed to find the optimal latent spaces. Finally, experiments on synthetic and real-world data show that our algorithms are able to preserve the network structures, and can be used for recommender systems and document clustering.

There are still some interesting further study points. For example, the dimension number of the discovered latent space needs to be predefined in the present algorithms. The ability to automatically find an optimized number for the shared latent space will make multi-level network factorization more practical for real-world tasks.

\section*{Acknowledgments}

This work was supported in part by the Australian Research Council (ARC) under discovery grant DP140101366 and the China Scholarship Council, and by the National Science Foundation of China under grant nos. 91024012, 91324005.

\ifCLASSOPTIONcaptionsoff
  \newpage
\fi

\bibliographystyle{IEEEtran}
\bibliography{TNN}

\begin{IEEEbiography}[{\includegraphics[width=1in,height=1.25in,clip,keepaspectratio]{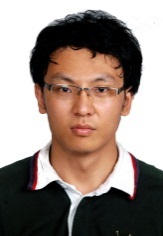}}]{Junyu Xuan}
is received the bachelor's degree in 2008 from China University of Geosciences, Beijing. Currently he is working toward the dual-doctoral degree both in Shanghai University and University of Technology, Sydney. His main research interests include Machine Learning, Complex Network and Web Mining.
\end{IEEEbiography}
\begin{IEEEbiography}[{\includegraphics[width=1in,height=1.25in,clip,keepaspectratio]{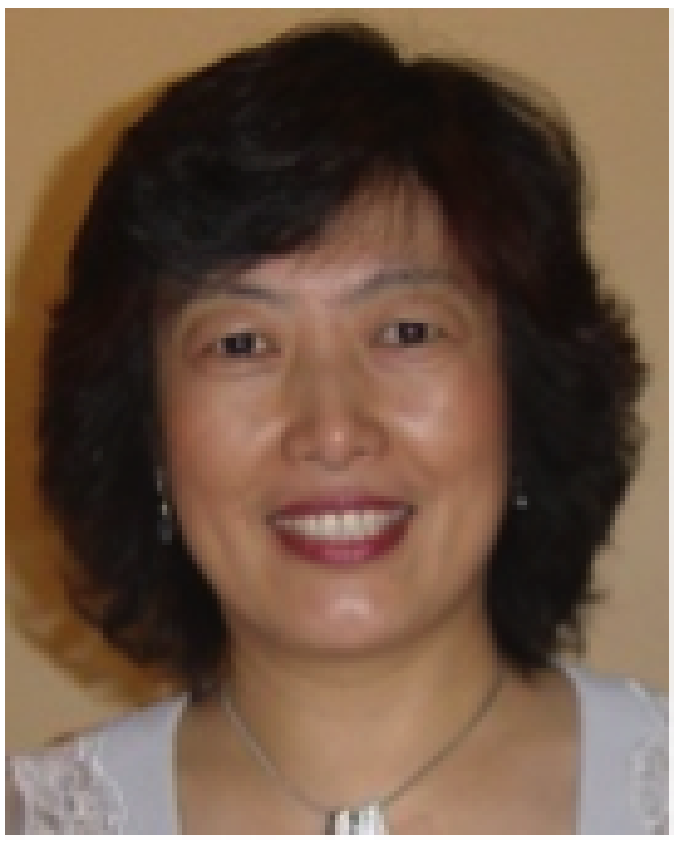}}]{Jie Lu}
is a full professor and Head of School of Software at the University of Technology, Sydney. Her research interests lie in the area of decision support systems and uncertain information processing. She has published five research books and 270 papers, won five Australian Research Council discovery grants and 10 other grants. She received a University Research Excellent Medal in 2010. She serves as Editor-In-Chief for Knowledge-Based Systems (Elsevier), Editor-In-Chief for International Journal of Computational Intelligence Systems (Atlantis), editor for book series on Intelligent Information Systems (World Scientific) and guest editor of six special issues for international journals, as well as delivered six keynote speeches at international conferences.
\end{IEEEbiography}
\begin{IEEEbiography}[{\includegraphics[width=1in,height=1.25in,clip,keepaspectratio]{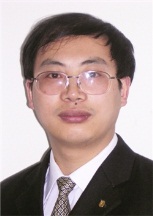}}]{Xiangfeng Luo}
is a professor in the School of Computers, Shanghai University, China. Currently, he is a visiting professor in Purdue University. He received the master's and PhD degrees from the Hefei University of Technology in 2000 and 2003, respectively. He was a postdoctoral researcher with the China Knowledge Grid Research Group, Institute of Computing Technology (ICT), Chinese Academy of Sciences (CAS), from 2003 to 2005. His main research interests include Web Wisdom, Cognitive Informatics, and Text Understanding. He has authored or co-authored more than 50 publications and his publications have appeared in IEEE Trans. on Automation Science and Engineering, IEEE Trans. on Systems, Man, and Cybernetics-Part C, IEEE Trans. on Learning Technology, Concurrency and Computation: Practice and Experience, and New Generation Computing, etc. He has served as the Guest Editor of ACM Transactions on Intelligent Systems and Technology. Dr. Luo has also served on the committees of a number of conferences/workshops, including Program Co-chair of ICWL 2010 (Shanghai), WISM 2012 (Chengdu), CTUW2011 (Sydney) and more than 40 PC members of conferences and workshops.
\end{IEEEbiography}
\begin{IEEEbiography}[{\includegraphics[width=1in,height=1.25in, clip]{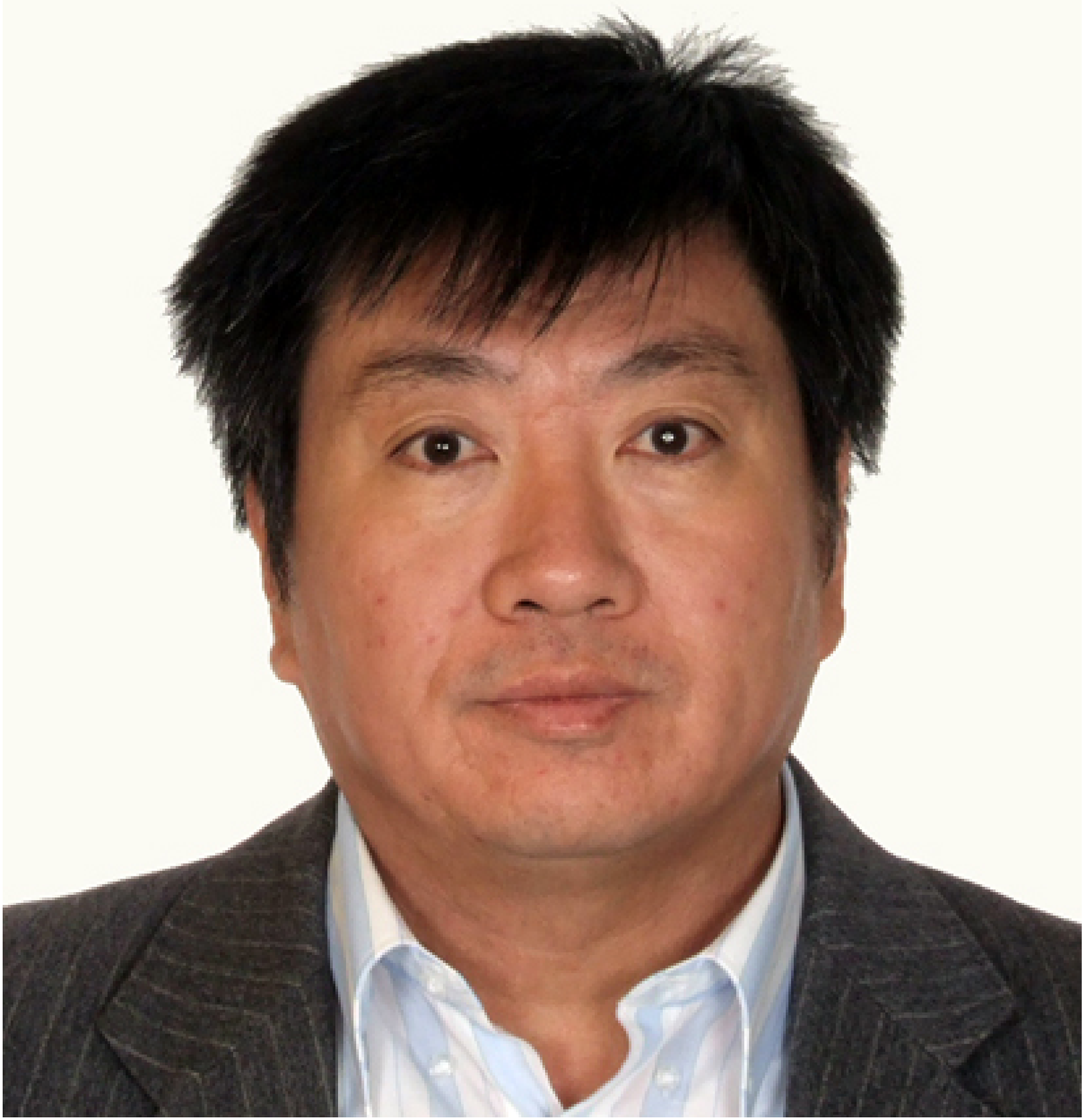}}]{Guangquan Zhang }
is an associate professor in Faculty of Engineering and Information Technology at the University of Technology Sydney (UTS), Australia. He has a PhD in Applied Mathematics from Curtin University of Technology, Australia. He was with the Department of Mathematics, Hebei University, China, from 1979 to 1997, as a Lecturer, Associate Professor and Professor. His main research interests lie in the area of multi-objective, bilevel and group decision making, decision support system tools, fuzzy measure, fuzzy optimization and uncertain information processing. He has published four monographs, four reference books and over 200 papers in refereed journals and conference proceedings and book chapters. He has won four Australian Research Council (ARC) discovery grants and many other research grants.
\end{IEEEbiography}

\end{document}